\documentclass[sigplan,screen,nonacm]{acmart}
\usepackage{asystem-tech-report}

\usepackage{amsmath}
\usepackage{array}
\usepackage{booktabs}
\usepackage{microtype}
\usepackage{xspace}
\newenvironment{RoutePackCompactDisplay}{%
  \setlength{\abovedisplayskip}{0pt plus .5pt}%
  \setlength{\belowdisplayskip}{0pt plus .5pt}%
  \setlength{\abovedisplayshortskip}{0pt plus .5pt}%
  \setlength{\belowdisplayshortskip}{0pt plus .5pt}%
  \setlength{\jot}{.5pt}%
}{}
\newenvironment{RoutePackDenseDisplay}{%
  \setlength{\abovedisplayskip}{0pt plus .5pt}%
  \setlength{\belowdisplayskip}{0pt plus .5pt}%
  \setlength{\abovedisplayshortskip}{0pt plus .5pt}%
  \setlength{\belowdisplayshortskip}{0pt plus .5pt}%
  \setlength{\jot}{.5pt}%
}{}
\newenvironment{RoutePackFixedRowDisplay}{%
  \setlength{\abovedisplayskip}{-1.5pt plus .5pt}%
  \setlength{\belowdisplayskip}{-1.5pt plus .5pt}%
  \setlength{\abovedisplayshortskip}{-1pt plus .5pt}%
  \setlength{\belowdisplayshortskip}{0pt plus .5pt}%
  \setlength{\jot}{.5pt}%
}{}

\newcounter{algorithm}
\newcounter{algoline}
\newenvironment{routealgorithm}[1]{%
  \par\noindent\begin{minipage}{\columnwidth}%
  \begingroup
  \refstepcounter{algorithm}%
  \setcounter{algoline}{0}%
  \setlength{\parindent}{0pt}%
  \setlength{\parskip}{0pt}%
  \footnotesize
  \vskip 0.8pt
  \hrule height 0.4pt\relax
  \vskip 2.5pt
  \noindent\textbf{Algorithm~\thealgorithm: #1}\par
  \vskip 2pt
  \hrule height 0.3pt\relax
  \vskip 2.5pt
  \ignorespaces
}{%
  \par
  \vskip 2.5pt
  \hrule height 0.4pt\relax
  \endgroup
  \end{minipage}\par
}
\newenvironment{routealgorithmfloat}[1]{%
  \begin{figure}[!t]
  \Description{Pseudocode for #1.}%
  \begin{routealgorithm}{#1}%
}{%
  \end{routealgorithm}
  \end{figure}%
}
\newcommand{\AlgLine}[1]{%
  \stepcounter{algoline}\noindent\makebox[1.8em][r]{\scriptsize\thealgoline:}%
  \hspace{0.5em}\parbox[t]{\dimexpr\linewidth-2.3em\relax}{\raggedright #1}\par
}
\newcommand{\AlgRequire}[1]{%
  \noindent\textbf{Require:} #1\par
}
\newcommand{\AlgEnsure}[1]{%
  \noindent\textbf{Ensure:} #1\par
}

\newcommand{\sysname}{RoutePack\xspace}
\newcommand{\RoutePackTableFont}{\small}

\title[RoutePack]{\sysname: Replay-Guided Expert Placement and Rollout Trajectory Scheduling for MoE Reinforcement Learning}
\reportauthor{Yibo Shen}{1}
\reportauthor{Xudong Han}{1}
\reportauthor{Xiaowei Zhu}{1}
\reportauthor{Gen Li}{1}
\reportauthor{Zhenxuan Pan}{1}
\reportaffiliation{1}{Ant Group}
\reportauthornotes{Corresponding author: \href{mailto:syb544241@antgroup.com}{syb544241@antgroup.com}}
\renewcommand{\shortauthors}{Shen et al.}

\begin{document}
\raggedbottom

\begin{abstract}
Mixture-of-Experts (MoE) reinforcement learning (RL) must balance two coupled
loads. The rollout trajectories in each data-parallel (DP) microbatch set its
dense-side work (attention and dense FFN), while token routing sets the expert
work of each expert-parallel (EP) rank. Balancing either load alone can worsen
the other: length-aware trajectory scheduling may concentrate expert demand,
while expert-aware scheduling may create dense-side stragglers. MoE RL makes
joint balancing possible: routing replay reveals each trajectory's token length
and layer-wise expert demand before training starts.

We present \sysname{}, a replay-guided planner that coordinates expert
placement with rollout-trajectory scheduling. Its key insight is that the two
controls act at different scales. Placement sets a window-level EP-rank load
bound that no trajectory schedule can cross; scheduling then removes the
hotspots that remain inside individual micro-steps under that placement.
\sysname{} first swaps experts across EP ranks incrementally to lower this
bound, keeping unmoved experts resident. With placement fixed, it reschedules
complete trajectories into token-capped microbatches, jointly balancing
dense-side and expert work at a fixed, capacity-feasible number of micro-steps
within the deployed synchronization groups. Placement stays fixed through
forward and backward execution, preserving logical routing and training-state
consistency without microbatch-level expert replication.

We evaluate mixed-task RL post-training of two hybrid-attention MoE models
(7.9B and 124B) in six NVIDIA H20-3e configurations spanning 8K--32K contexts,
with cross-GPU checks of the 7.9B model on H200 and H20. In trace replay,
incremental placement meets the load-balance target while preserving expert
residency, and rescheduling further reduces modeled per-micro-step deficits.
Measured over the Megatron training region, with planning and state loading
reported separately, expert placement alone raises tokens per second (TPS) by
2.67--6.24\% over an in-house DP-local first-fit-decreasing (FFD) packing
baseline, and full \sysname{} raises it by 4.44--11.85\%. Full \sysname{} also
leads on both additional GPUs.
\end{abstract}

\begin{teaserfigure}
  \centering
  \includegraphics[width=\textwidth]{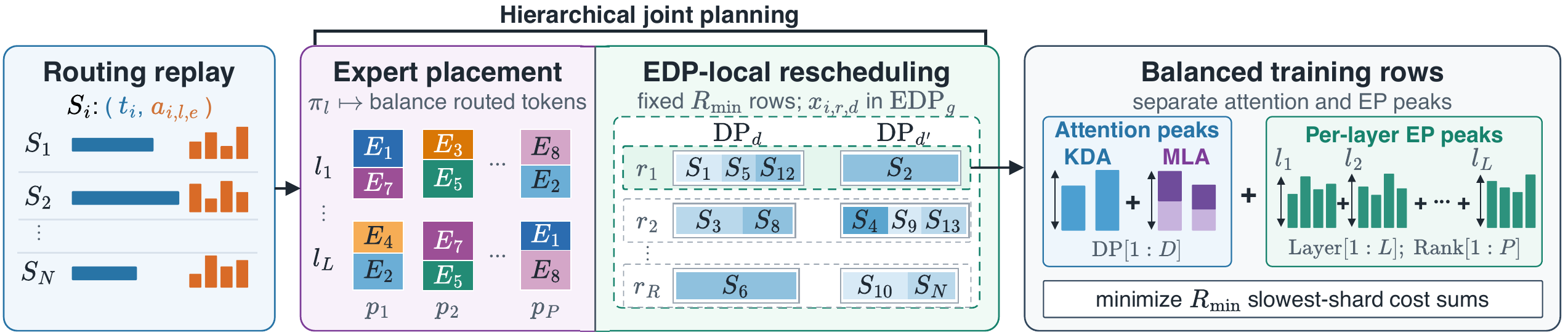}
  \Description{\sysname{} replays each trajectory's token length and layer-wise
  expert demand. A hierarchical planner couples per-layer expert placement
  with fixed-row trajectory schedules scored over the runtime's EDP shards. The resulting
  training rows reduce separate KDA, MLA, and per-layer expert-parallel peak
  costs.}
  \caption{The general workflow of \sysname: replay-guided expert placement and
  EDP-aware fixed-row trajectory rescheduling.}
  \label{fig:routepack-overview}
\end{teaserfigure}

\maketitle
\hypersetup{pdfauthor={Yibo Shen, Xudong Han, Xiaowei Zhu, Gen Li, Zhenxuan Pan}}

\section{Introduction}
\label{sec:introduction}

Sparse Mixture-of-Experts (MoE) models increase capacity while activating only
a small subset of feed-forward experts for each token. Their training steps
nevertheless expose two distinct load tails. Dense-side computation executes
independently within data-parallel (DP) replicas and includes every attention
sublayer and dense FFN, so its critical path depends on the lengths and
composition of the rollout trajectories in each microbatch. Expert computation
is distributed through expert parallelism (EP), where grouped GEMMs are paced
by the physical rank receiving the most routed tokens. Dispatch and combine
also depend on the physical routing layout. RL post-training amplifies both
effects: rollout trajectories vary substantially in length, while expert popularity
changes across optimizer windows~\cite{deepseekv3,relibra,foremoe}. In our
replay corpus under identity placement, the hottest expert receives up to
$10.2\times$ its uniform share, and the busiest EP rank of a layer carries up to
$2.35\times$ the mean rank load (Section~\ref{sec:pilot-rlvr-skew}).

Existing systems address one side at a time. Expert-side systems optimize
transport, placement, replication, or post-route assignment
~\cite{deepep,relibra,foremoe,ultraep,moonep,finemoe}. RL frameworks pack
complete responses for capacity~\cite{verlpacking,openrlhfpacking,nemorlpacking},
while schedulers reorder them by length~\cite{libra,harts}. Neither jointly
selects physical expert placement and the capacity-constrained trajectory
schedule, and optimizing one side alone can worsen the other: on Tiny, expert
placement alone raises the modeled row-local EP excess from 9.40\% to 23.17\%
of uniform EP work (Section~\ref{sec:mechanism-decomposition}).

MoE RL provides an unusual opportunity to coordinate the two decisions. The
rollout and training phases consume the same generated rollout trajectories.
Routing replay can therefore expose each trajectory's token length and
layer-wise expert demand before a \emph{sealed optimizer window}---the complete
rollout batch consumed by one optimizer update---is admitted to training
~\cite{r3,relibra}. This turns data layout into a load-balancing control:
Figure~\ref{fig:routepack-overview} coordinates layer-wise expert placement
with an expert-data-parallel (EDP)-aware trajectory layout for the same window.

We present \sysname, a replay-guided hierarchical planner that coordinates
layer-wise expert placement with rollout-trajectory rescheduling. The two
controls operate at different scales. \sysname{} first uses aggregate window
demand to lower the EP-rank straggler bound that trajectory reordering cannot
cross, swapping experts incrementally so that unmoved experts stay resident.
With that placement fixed, it assigns the complete trajectory set to
\emph{execution rows}: synchronized micro-steps in which every DP rank
processes one token-capped microbatch. It fixes the smallest
capacity-feasible row count it can certify, or the best-known one, before
balancing load. It then scores each row by the slowest DP microbatch of each dense-side stage plus the
busiest EP rank of each MoE layer, and minimizes the total row cost of the
slowest EDP shard.
Sections~\ref{sec:formulation} and~\ref{sec:solver} give the objective and its
capacity-feasible population-annealing solver, respectively.

We integrate \sysname{} into an in-house post-training framework. After rollout
seals a window, replay profiles drive CPU-side placement repair and trajectory
rescheduling while post-rollout processing and state preparation proceed in
parallel. Before training admission, the framework commits the expert layout
and execution rows. Expert parameters and their expert-indexed training state move into
pre-existing physical slots, so tensor shapes, gradient buffers, communicators,
and the training unit remain unchanged.

This paper makes three contributions:
\begin{list}{\labelitemi}{%
  \setlength{\leftmargin}{\leftmargini}%
  \setlength{\labelsep}{0.5em}%
  \setlength{\labelwidth}{\dimexpr\leftmargin-\labelsep\relax}%
  \setlength{\topsep}{3pt}%
  \setlength{\partopsep}{0pt}%
  \setlength{\itemsep}{1pt}%
  \setlength{\parsep}{0pt}%
}
  \item We show that expert placement and trajectory scheduling control
  different scales of MoE imbalance: placement sets a window-level EP-rank
  bound that scheduling cannot cross, while scheduling sets row-local peaks
  under that placement. Replay-guided hierarchical planning coordinates both
  for whole trajectories in MoE RL.

  \item We formulate an EDP-aware objective that first fixes a
  capacity-efficient row count, then jointly minimizes dense-side work and
  physical EP-rank tails.

  \item We implement state-consistent incremental placement and parallel
  population annealing, overlapping planning with state loading (97.4--99.4\%
  median overlap). Across six configurations, full \sysname{} records
  4.44--11.85\% higher post-admission TPS in the Megatron training region than
  the in-house DP-local FFD packing baseline. In a single-window Tiny 16K device trace, it
  lowers DeepEP dispatch and combine waits by 66.0\% while total compute
  changes by only 0.12\%.
\end{list}

\section{Background and Motivation}
\label{sec:background}

\sysname{} couples trajectories and DP-slot schedule cells, logical experts and
physical EP ranks, and the EDP shards that define synchronization domains.
Table~\ref{tab:notation} groups the notation by scope; for any positive integer
$K$, we write $[K]=\{1,\ldots,K\}$.

\begin{table}[!t]
  \caption{Core notation and index scopes.}
  \label{tab:notation}
  \centering
  \scriptsize
  \renewcommand{\arraystretch}{1.10}
  \setlength{\tabcolsep}{2pt}
  \begin{tabular}{@{}>{\centering\arraybackslash}p{0.15\columnwidth}
                      @{\hspace{4pt}}
                      >{\raggedright\arraybackslash}p{0.84\columnwidth}@{}}
    \toprule
    Symbol & Scope and meaning \\
    \midrule
    \multicolumn{2}{@{}l}{\textit{Indices and sets}} \\
    $i$ & Trajectory index in a sealed optimizer window, $i\in[N]$. \\
    $r$ & Execution-row index, $r\in[R]$; $R$ is the row count. \\
    $d$ & Controller DP-slot index, $d\in[D]$; each row has $D$ cells. \\
    $g$ & EDP-shard index, $g\in\mathcal G$; unrelated shards do not synchronize. \\
    $l$ & MoE-layer index, $l\in\mathcal L_{\mathrm{MoE}}$; dense layers are excluded. \\
    $e$ & Logical expert index, $e\in\mathcal E_l$, in MoE layer $l$. \\
    $p$ & Physical EP-rank index, $p\in[P]$, in the relevant communicator. \\
    $s$ & Dense-side stage type (attention or dense FFN), $s\in\mathcal A$. \\
    \midrule
    \multicolumn{2}{@{}l}{\textit{Replay input and topology}} \\
    $S_i,\ t_i$ & Trajectory $i$ and its token length. \\
    $B_{r,d}$ & Schedule cell at $(r, d)$; its assigned trajectories form a microbatch. \\
    $\mathbf A_i,\ a_{i,l,e}$ & Trajectory routing matrix and layer--expert token-load entry. \\
    $\mathcal D_g$ & DP slots projected onto EDP shard $g$. \\
    $\pi$ & Layer-wise placement plan; $\pi_l(e)$ is expert $e$'s physical slot and $\operatorname{owner}(\pi_l(e))$ its EP rank. \\
    \midrule
    \multicolumn{2}{@{}l}{\textit{Layout, load, objective, and runtime}} \\
    $C$ & Token capacity of one cell and its resident microbatch. \\
    $q_{i,l,p}$ & Trajectory $i$ demand on rank $p$ at layer $l$ after placement. \\
    $L_{l,e}$ & Window-total demand of expert $e$ at layer $l$. \\
    $G_{l,p}$ & Window-total demand on rank $p$ at layer $l$. \\
    $\alpha,\ \beta$ & Token-linear weight $\alpha_s$ and token-pair weight $\beta_s$ of dense-side stage $s$. \\
    $W_{r,g,l,p}$ & Row $r$, shard $g$, layer $l$ load on EP rank $p$. \\
    $\mathcal X_R$ & Capacity-feasible layouts containing exactly $R$ rows. \\
    $R_{\min},\ R^\star$ & Minimum feasible and planner-returned row counts. \\
    $x$ & Complete layout; $x_{i,r,d}$ is the assignment of trajectory $i$ to cell $(r,d)$, and $x^\star$ is the returned layout.\\
    \bottomrule
  \end{tabular}
\end{table}

\begin{figure}[!t]
  \centering
  \includegraphics[width=\columnwidth]{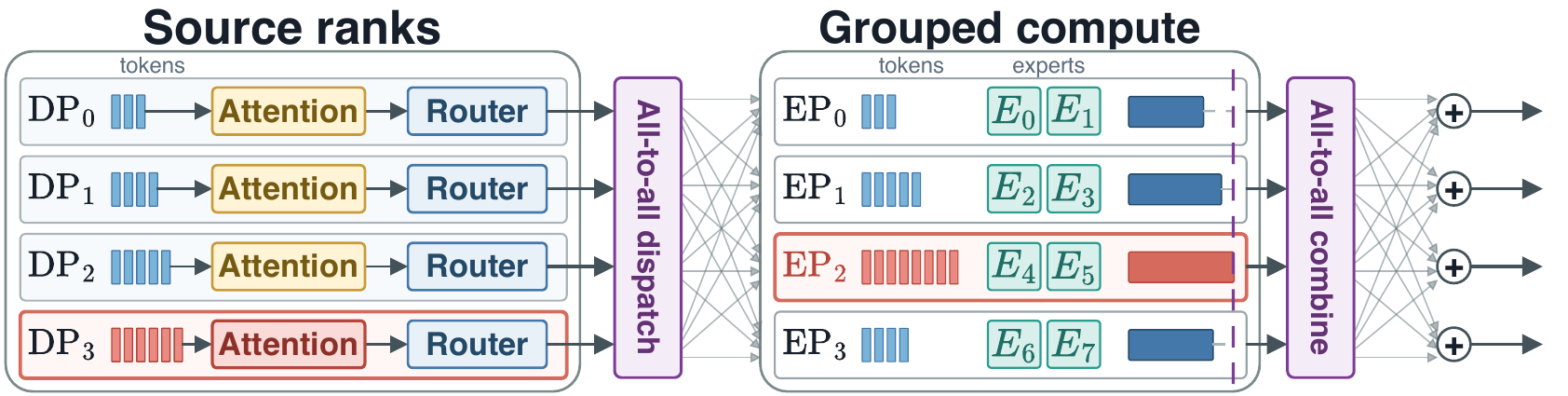}
  \Description{Four data-parallel lanes run attention and routing, then dispatch
  tokens through all-to-all communication to four expert-parallel ranks. Each
  rank contains two local experts and grouped compute. The busiest expert rank
  is highlighted, followed by all-to-all combine and residual additions.}
  \caption{ MoE layer and expert parallelism. The straggler is defined by the highlighted DP and EP ranks.}
  \label{fig:moe-ep-execution}
\end{figure}

\subsection{MoE RL Training with Routing Replay}
\label{sec:background-workflow}

In decoupled RL systems, rollout forms a sealed optimizer window that one
optimizer update later consumes. Before training, systems pack variable-length
trajectories into token-capped microbatches on data-parallel (DP)
workers~\cite{verlpacking,openrlhfpacking,
nemorlpacking}. We use \emph{cell} for the schedule position $B_{r,d}$; the
trajectories assigned to that cell form its resident microbatch. \sysname{}
organizes $RD$ such cells into $R$ aligned execution rows, with one cell per DP
slot in each row. An execution row is a physical scheduling unit; it is distinct from an
algorithmic minibatch, which defines the learning update rather than
the synchronized device layout.

With rollout routing replay (R3), rollout records token-level logical top-$k$
expert IDs and training replays the same IDs~\cite{r3,relibra}. We aggregate these IDs into
$\mathbf A_i$ before planning; placement changes only their physical owners,
not the logical token-to-expert assignments. \sysname{} therefore receives

\begin{RoutePackCompactDisplay}
\begin{equation}
  \{(t_i,\mathbf A_i)\}_{i=1}^{N},\qquad
  \mathbf A_i=[a_{i,l,e}]_{l\in\mathcal L_{\mathrm{MoE}},e\in\mathcal E_l},
  \label{eq:routing-replay-input}
\end{equation}
\end{RoutePackCompactDisplay}

for every such window. Appendix~A describes how the
replayed IDs are aggregated and gives the
operator-level interpretation of the dense-side model used below.

Figure~\ref{fig:moe-ep-execution} locates the two parallel scopes. Dense-side
operators and routing run locally in each DP cell $B_{r,d}$; routed assignments
are then aggregated across the cells in an EDP shard $\mathcal D_g$, and each
EP owner groups its dispatched tokens.

\subsection{Coupled Dense-Side and Expert Stragglers}
\label{sec:background-imbalance}

RL responses vary substantially in length, so equal trajectory counts need not
produce equal dense-side work. Dense-side Transformer operators fall into two
relevant complexity classes. Dense FFNs, token-wise projections, recurrent or
linear attention such as Kimi Delta Attention (KDA)~\cite{kimilinear}, and
fixed-width sliding-window attention scale with the total tokens in a
microbatch, $\sum_{i\in B}t_i$. Full attention instead forms pairwise
interactions within each trajectory and scales with $\sum_{i\in B}t_i^2$;
Multi-Query Attention (MQA), Grouped-Query Attention (GQA), and Multi-Head
Latent Attention (MLA) reduce projection or KV-state constants but retain this
token-pair term~\cite{transformer,gqa,deepseekv2}.
Hybrid architectures interleave these two classes~\cite{kimilinear,qwen3next}.
We therefore approximate the cost of dense-side stage $s$ on cell $B$ by

\begin{RoutePackCompactDisplay}
\begin{equation}
  F_s(B)=\frac{\alpha_s}{C}\sum_{i\in B}t_i+
         \beta_s\sum_{i\in B}(t_i/C)^2,\qquad \beta_s\ge0.
  \label{eq:attention-cost-proxy}
\end{equation}
\end{RoutePackCompactDisplay}

The linear term captures dense FFNs, projections, and recurrent work, while the
quadratic term captures pairwise attention. In our evaluated models, KDA and
dense FFNs use $\beta_s=0$ and MLA uses $\beta_s>0$; operator- and
hardware-specific effects are absorbed into stage-calibrated coefficients. This
compact proxy is sufficient for the joint planning formulation; its
kernel-specific limits are evaluated separately.

Expert-side work depends on both the trajectory and layer. After applying
$\pi_l$, an
EDP shard aggregates the routed tokens assigned to each physical EP rank, and
dispatch plus grouped expert computation is paced by the busiest rank at that
layer. An EP rank may own several experts, but grouped-GEMM backends execute
their local MLPs as one grouped operation. We therefore first sum demand over
the experts owned by a rank and then take the maximum across synchronizing EP
ranks, matching the abstraction used by exact-load systems such as
UltraEP~\cite{ultraep}.

\label{sec:background-insufficient}
The two load tails require complementary controls. Under fixed placement
$\pi_l$, rescheduling leaves each physical rank's window-total demand
unchanged:

\begin{RoutePackCompactDisplay}
\begin{equation}
  G_{l,p}(\pi_l)
  =\sum_i\sum_{e:\operatorname{owner}(\pi_l(e))=p}a_{i,l,e}.
  \label{eq:aggregate-invariance}
\end{equation}
\end{RoutePackCompactDisplay}

Consequently, every feasible schedule $x$ obeys

\begin{RoutePackCompactDisplay}
\begin{equation}
  \sum_r\sum_g\max_p W_{r,g,l,p}(\pi,x)
  \ge \max_p G_{l,p}(\pi_l).
  \label{eq:packing-lower-bound}
\end{equation}
\end{RoutePackCompactDisplay}

Trajectory rescheduling alone cannot cross this placement-defined lower bound. Conversely,
equal window-total demands do not imply balanced rows: correlated routing
vectors can expose the same physical rank together. Moving a whole trajectory also
changes its linear and quadratic dense-side terms together with one routing
vector per MoE layer. Length-only scheduling can therefore create an expert-side
hotspot, while expert-only layout optimization can create a dense-side outlier;
conventional first-fit decreasing (FFD) cannot distinguish equal-length trajectories with different
routing signatures.

\begin{figure}[!t]
  \centering
  \includegraphics[width=\columnwidth]
    {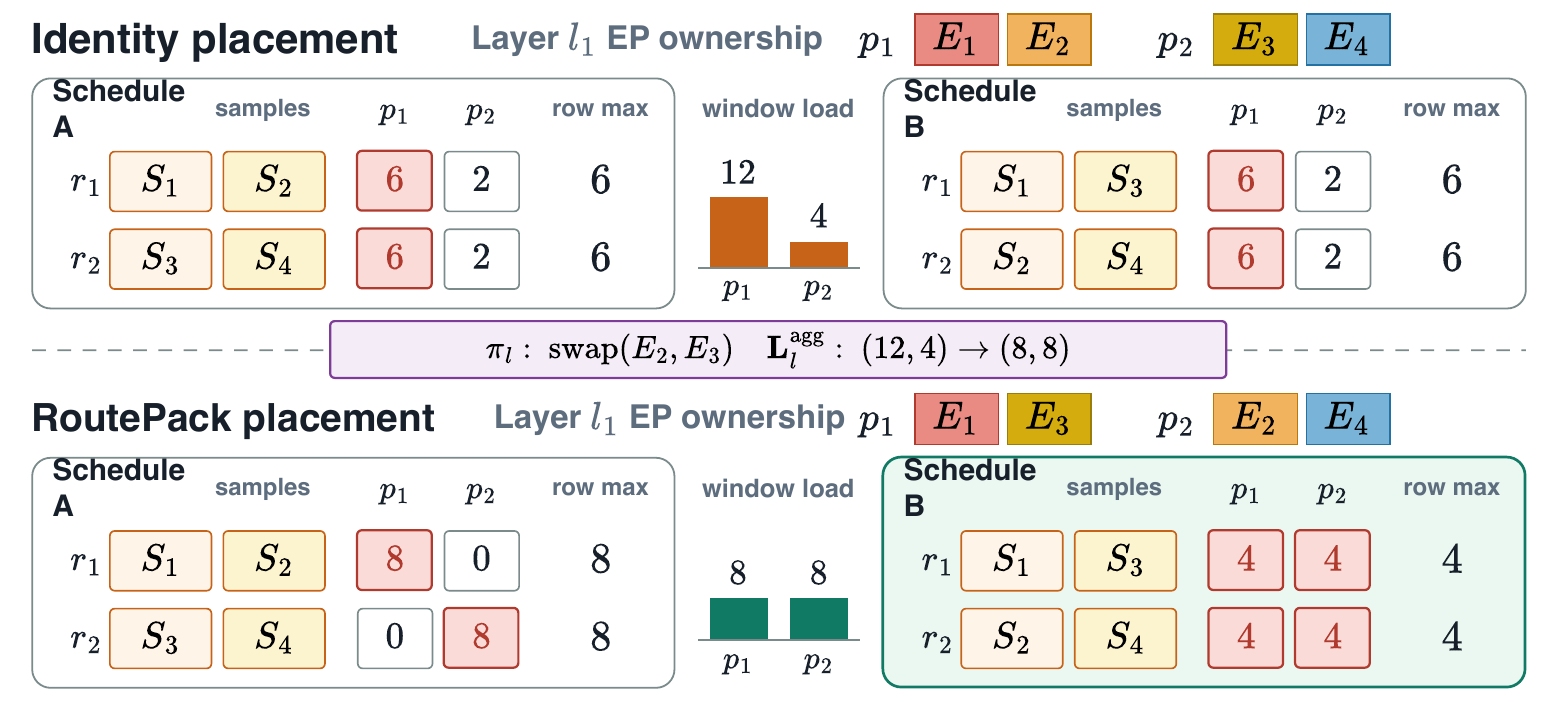}
  \Description{Under identity expert placement, two trajectory schedules retain a
  twelve-versus-four window-total rank load and a row peak of six. A
  layer-local expert permutation balances the window load at eight versus
  eight. Routing-aware rescheduling then lowers both row peaks from eight to four
  without adding rows.}
  \caption{Placement lowers aggregate skew; rescheduling reduces row-local peaks
  at fixed row count.}
  \label{fig:placement-packing}
\end{figure}

Figure~\ref{fig:placement-packing} illustrates the resulting hierarchical order.

\subsection{Pilot Studies and Replay Benchmark}
\label{sec:background-pilot}

A grouped-GEMM microbenchmark tests whether rank-total routed tokens predict
grouped-expert time; a 20-window routing corpus is the common benchmark for the
later offline placement, rescheduling, solver, and communication studies.

\subsubsection{Validating the EP-Compute Proxy}
\label{sec:pilot-ep-proxy}

\sysname{} approximates a row's expert-compute time by its largest
post-dispatch rank-total token count. We test this proxy independently of the
mixed-task corpus with two Flash grouped-GEMM controls. A \emph{rank-total
sweep} measures 7,680 naturally occurring shapes from 480 layer--EP-group
observations (16 ranks per group and 32 local experts per rank) after five
warmup and 20 timed bf16 forward--backward iterations. A \emph{fixed-total
shape control} redistributes 81,830 tokens from one measured rank shape nearly
uniformly across its local experts. The former tests whether rank-total demand
predicts latency; the latter isolates residual intra-rank shape effects. For a
positive-mean vector $\mathbf z=(z_1,\ldots,z_m)$, we report the population
coefficient of variation:
\begin{RoutePackCompactDisplay}
\begin{equation}
  \mathrm{CV}(\mathbf z)=\frac{\sqrt{\frac{1}{m}\sum_j(z_j-\bar z)^2}}{\bar z}
\end{equation}
\end{RoutePackCompactDisplay}
Expert Max/Avg and CV here range over local experts; Section~\ref{sec:expert-placement}
applies the same statistics to physical EP-rank loads.

\begin{center}
\begin{minipage}{\columnwidth}
  \centering
  \begingroup
    \captionsetup{hypcap=false}
    \captionof{table}{Fixed-token rank-local control at 81,830 tokens.}
    \label{tab:rank-local-shape-control}
  \endgroup
  \RoutePackTableFont
  \setlength{\tabcolsep}{3pt}
  \begin{tabular}{lccc}
    \toprule
    Shape & Expert Max/Avg & Expert CV & P50 latency (ms) \\
    \midrule
    Real & 8.151 & 1.848 & 6.8772 \\
    Uniform & 1.000 & 0.000 & 6.6157 \\
    \bottomrule
  \end{tabular}
  \par\kern 4pt
  \includegraphics[width=0.97\columnwidth]{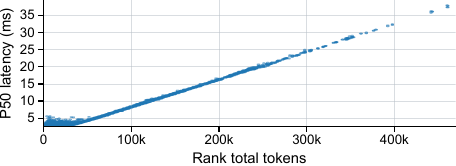}
  \Description{Scatter plot of grouped-GEMM forward-and-backward median latency
  against total tokens on one expert-parallel rank.}
  \begingroup
    \captionsetup{hypcap=false}
    \captionof{figure}{Grouped-GEMM forward--backward P50 latency versus
    rank token count (7,680 shapes).}
    \label{fig:grouped-gemm-cost-model}
  \endgroup
\end{minipage}
\end{center}

Above the launch-latency floor, grouped-GEMM P50 latency grows approximately
with rank total tokens (Figure~\ref{fig:grouped-gemm-cost-model}); token
max/mean has P50/P90 values of 3.5244/5.0788, versus 2.8099/3.9500 for latency
max/mean.
At fixed rank total, uniformizing the local-expert shape changes P50 latency by
only 3.8\% (Table~\ref{tab:rank-local-shape-control}). Aggregate rank demand is
therefore the dominant expert-compute proxy in this regime. The sweep measures
grouped-GEMM compute; Appendix~C gives the complete protocol.

\subsubsection{Shared Offline Replay Corpus}
\label{sec:pilot-rlvr-skew}
\label{sec:offline-replay-corpus}

We collect 20 consecutive sealed
optimizer windows from each of Ling-3.0-Tiny (Tiny) and Ling-3.0-Flash
(Flash)~\cite{ling3tiny,ling3models,ling3flash}. Tiny has 7.9B total and 1.3B
activated parameters per token, 23 MoE layers, and 128 experts per layer;
Flash has 124B total and 5.1B activated parameters per token, 40 MoE layers,
and 512 experts per layer. Both hybrid-attention models combine KDA and MLA
attention layers.

Both runs use an in-house dataset comprising math, code, STEM, and logic tasks.
Each window contains 64 prompt groups with eight GRPO rollouts per
prompt~\cite{deepseekmath}, hence 512 complete rollout trajectories, each
capped at 8,192 tokens. We preserve the task labels and the realized sampling
mixture.

Table~\ref{tab:offline-corpus} summarizes the retained workload (10,240
trajectories per model). The different length
distributions are useful rather than normalized away: Tiny stresses nearly
capacity-sized cells, while Flash contains a broader mix of short and long
trajectories. The task-mix column reports math/code/STEM/logic percentages in
that order.

\begin{table}[H]
  \centering
  \caption{Shared offline replay corpus (20 consecutive windows per model).}
  \label{tab:offline-corpus}
  \RoutePackTableFont
  \setlength{\tabcolsep}{2.5pt}
  \begin{tabular*}{\columnwidth}{@{\extracolsep{\fill}}lrrr@{}}
    \toprule
    Model & Count tensor/window & Length P50/P95 & Task mix (\%) \\
    \midrule
    Tiny  & $512\!\times\!23\!\times\!128$ & 7,298/7,890
          & 58.7/15.2/22.1/4.1 \\
    Flash & $512\!\times\!40\!\times\!512$ & 3,178/7,323
          & 56.4/15.5/23.6/4.5 \\
    \bottomrule
  \end{tabular*}
\end{table}

\begin{figure}[t]
  \centering
  \includegraphics[width=\columnwidth]
    {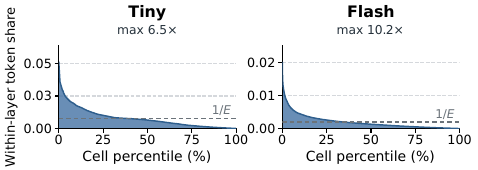}
  \par\vspace{-1.0mm}
  {\footnotesize\textbf{(a)}}\par\vspace{-0.6mm}
  \includegraphics[width=\columnwidth]
    {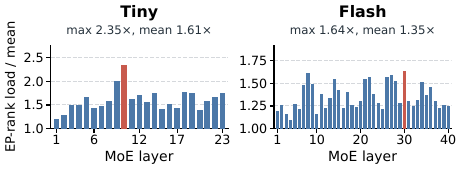}
  \par\vspace{-1.0mm}
  {\footnotesize\textbf{(b)}}\par\vspace{-0.6mm}
  \includegraphics[width=\columnwidth]
    {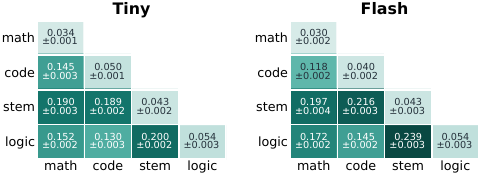}
  \par\vspace{-1.0mm}
  {\footnotesize\textbf{(c)}}
  \Description{Pilot routing statistics for Tiny and Flash: pooled
  layer-expert shares, physical-rank max-to-mean loads by layer, and
  category-pair Jensen-Shannon distances.}
  \caption{Routing skew under identity EP=8: (a) pooled within-layer expert
  shares (dashed: $1/E$); (b) rank max/mean per layer; (c) category-pair
  square-root Jensen--Shannon divergence ($\sqrt{\mathrm{JSD}}$).}
  \label{fig:pilot-routing-skew}
\end{figure}

We characterize expert-, rank-, and window-category imbalance under identity
EP=8, before either RoutePack control (Figure~\ref{fig:pilot-routing-skew}).
Panels (a) and (b) pool all 20 windows. Panel (c) normalizes each
window--category rank profile before computing distances; off-diagonal cells
average cross-category window pairs, while diagonal cells compare windows of
the same category. Windows are bootstrap blocks.

\begin{table*}[!t]
  \centering
  \caption{Signals and controls used by related load balancers.}
  \label{tab:related-controls}
  \resizebox{\textwidth}{!}{%
  \begin{tabular}{@{}llllll@{}}
    \toprule
    Method & Planning signal & Scope & Expert control & Data control & Target \\
    \midrule
    RL Frameworks~\cite{areal,slimepacking,verlpacking}
      & Length & RL pre-step & -- & Token-capped bins & Capacity/length \\
    Libra~\cite{libra}
      & Length + attention cost & LLM training & -- & Trajectory/tile scheduling & Attention balance \\
    Semantic Parallelism~\cite{semanticparallelism}
      & Co-activation & MoE serving & Expert placement & Request/token reshuffling & EP locality \\
    Gimbal~\cite{gimbal}
      & Source pressure & MoE serving & Expert placement & Request admission & Load/comm./migration \\
    ReLibra~\cite{relibra}
      & Routing replay & MoE RL & Reordering + replicas & Source assignment + token split & Compute/communication \\
    DeepEP~\cite{deepep}
      & Routed traffic & Train/infer & Transport only & Post-route transport & Dispatch/combine \\
    UltraEP~\cite{ultraep}
      & Live routing & RSN layer/step & Dynamic replicas & Replica-aware token split & Load/communication \\
    MoonEP~\cite{moonep}
      & Live routing & NVLink layer/step & Replicas + prefetch & Replica-aware token split & Balance/locality \\
    FineMoE~\cite{finemoe}
      & Routed load & MoE training & Replica placement & LP token scheduling & EP compute balance \\
    ForeMoE~\cite{foremoe}
      & Routing foresight & MoE RL micro-step & Relocate + replicate & Post-route assignment & Compute/transfer \\
    \textbf{\sysname}
      & \textbf{Length + route replay} & \textbf{Sealed optimizer window}
      & \textbf{Accumulative relocation} & \textbf{Trajectory scheduling}
      & \textbf{Dense-side + EP tails} \\
    \bottomrule
  \end{tabular}%
  }
\end{table*}

Figure~\ref{fig:pilot-routing-skew} shows that RL post-training is imbalanced
at two nested granularities. The hottest layer--expert cell receives
$6.5\times$ and $10.2\times$ the uniform share for Tiny and Flash, respectively
(a). Equal ownership still leaves worst-layer EP-rank max/mean ratios of
$2.35\times$ and $1.64\times$ (b). Flash's 64 experts per rank smooth its
heavier expert tail more than Tiny's 16, showing that expert- and rank-level
imbalance are related but not interchangeable. Routing profiles are also
task-conditioned: within-category cross-window distances are only
$0.03$--$0.05$, whereas cross-category distances reach $0.12$--$0.24$
(c)~\cite{endres2003metric}. Thus, task families induce distinct and repeatable
physical-rank demand patterns.

Together, these observations motivate two
complementary controls. Placement reduces aggregate rank skew over the
optimizer window. Given that placement, a mixed window still contains
trajectories with heterogeneous, potentially complementary signatures;
rescheduling them across synchronized rows can reduce local rank peaks without
changing the window's sample multiset or aggregate routing demand.

\subsection{Related Work}
\label{sec:background-related}

Dynamic expert placement predates MoE RL: FasterMoE shadows popular
experts~\cite{fastermoe}, SmartMoE combines offline strategy search with online
selection~\cite{smartmoe}, FlexMoE remaps experts to devices from real-time
traffic~\cite{flexmoe}, and Lina prioritizes and balances all-to-all
traffic~\cite{lina}; all adapt to live or recently observed load.
Replay-guided expert systems are closest to ours in signal, but differ in
control granularity. ReLibra combines batch-level expert reordering and
source-GPU assignment with intra-node replication and token splitting at
microbatch granularity~\cite{relibra}. ForeMoE uses rollout foresight to
relocate or replicate experts and reassign routed tokens at each
micro-step~\cite{foremoe}. LAER-MoE couples expert re-layout with token-routing
decisions; Pro-Prophet plans balancing strategies and operator
schedules~\cite{laermoe,proprophet}. UltraEP, MoonEP, and FineMoE balance live
routed work using redundancy or replica-aware token scheduling, while DeepEP
optimizes dispatch and combine~\cite{ultraep,moonep,finemoe,deepep}. In
contrast, \sysname{} commits a state-consistent expert map for the training
interval and reschedules only complete trajectories.

Model--data co-scheduling also appears in MoE serving. Semantic Parallelism
collocates likely co-activated experts and requests, while Gimbal coordinates
expert placement and request admission~\cite{semanticparallelism,gimbal}.
Their target is inference locality rather than optimizer-window training.

Trajectory-side controls start from a different unit. RL frameworks construct
token-capped microbatches for capacity utilization and length
balance~\cite{areal,slimepacking,verlpacking,openrlhfpacking,nemorlpacking}.
Libra balances attention work; RollPacker groups long-tail rollouts into
selected generation rounds; HARTS co-plans prefix-compressed microbatches,
DP assignments, and slots for hybrid-attention RL~\cite{libra,rollpacker,harts}.
These systems do not couple each trajectory's layer-wise routing demand to
expert placement in fixed, capacity-feasible training rows.
Table~\ref{tab:related-controls} summarizes the distinct signals and controls.

\section{Hierarchical Planning and Expert Placement}
\label{sec:formulation}

\subsection{Fixed-Row Feasibility}
\label{sec:fixed-row}

Under fixed placement $\pi_l$, trajectory $S_i$ contributes the following demand
to physical EP rank $p$ at layer $l$:

\begin{RoutePackFixedRowDisplay}
\begin{equation}
  q_{i,l,p}(\pi_l)=
  \sum_{e:\operatorname{owner}(\pi_l(e))=p}a_{i,l,e}.
  \label{eq:physical-rank-demand}
\end{equation}
\end{RoutePackFixedRowDisplay}

Binary variables $x_{i,r,d}\in\{0,1\}$ assign each trajectory exactly once,
with every cell nonempty and within its token cap:

\begin{RoutePackFixedRowDisplay}
\begin{align}
  &\sum_{r=1}^{R}\sum_{d=1}^{D}x_{i,r,d}=1,
    &&\forall i, \label{eq:trajectory-bijection}\\
  &1\le\sum_i x_{i,r,d},
    &&\forall r,d, \label{eq:nonempty-cell}\\
  &\sum_i t_i x_{i,r,d}\le C,
    &&\forall r,d. \label{eq:token-capacity}
\end{align}
\end{RoutePackFixedRowDisplay}

These constraints define $\mathcal X_R$ for the execution rows introduced in
Section~\ref{sec:background-workflow}. At fixed topology and capacity, a smaller feasible $R$ raises the average token payload per cell, exposing
larger dense and grouped-expert kernels while amortizing launch, dispatch, and
synchronization overhead over more useful work. It also creates fewer row
boundaries at which participating ranks must synchronize, reducing execution fragmentation.
We therefore choose the row count before comparing load-balanced layouts, and hold it fixed during
that comparison. To maintain 1F1B pipeline occupancy, deployed search imposes at least $2\,\mathrm{PP}$ rows.
When fixed-bin feasibility is certified, the selected count is

\begin{RoutePackFixedRowDisplay}
\begin{equation}
  R^\star=\max\{R_{\min},2\,\mathrm{PP}\},\quad
  R_{\min}=\min\{R:\mathcal X_R\ne\varnothing\}.
  \label{eq:min-row-count}
\end{equation}
\end{RoutePackFixedRowDisplay}

Under a planning limit, $R^\star$ is instead the best-known feasible count
respecting the same pipeline floor. Every subsequent
search remains inside $\mathcal X_{R^\star}$. Section~\ref{sec:mechanism-decomposition}
reports row counts and workload gaps; Appendix~K separates
compaction from refinement at fixed $R^\star$.

With the peak objective below, fixed-row assignment is a 0--1 ILP after
standard epigraph linearization. Its variables scale with trajectory--cell pairs
and row--shard--layer--rank tuples, motivating the bounded specialized search
in Section~\ref{sec:solver}; Section~\ref{sec:ilp-comparison} assesses
generic-solver scaling.

\subsection{Layer-Wise Expert Placement}
\label{sec:expert-placement}

By Equation~\ref{eq:aggregate-invariance}, \sysname{} first forms
$L_{l,e}=\sum_i a_{i,l,e}$ independently at each MoE layer. Let
$\pi_l^{-}$ be the placement carried from the preceding window; the
cold start uses identity placement. For a candidate placement,

\begin{RoutePackFixedRowDisplay}
\begin{equation}
  G_{l,p}(\pi_l)=
  \sum_{e:\operatorname{owner}(\pi_l(e))=p}L_{l,e}.
  \label{eq:placement-load}
\end{equation}
\end{RoutePackFixedRowDisplay}

Starting from $\pi_l^{-}$, the planner repeatedly applies the cross-rank expert
swap with the largest strict reduction in $\sum_pG_{l,p}^2$. It stops when the
rank-load CV reaches $\epsilon=5\times10^{-4}$ or no improving swap remains.
This greedy swap repair, in the spirit of LPT balancing~\cite{graham1969lpt}, preserves the fixed number of experts per
rank and stops moving experts once the layer reaches the CV target.
Experts that stay on a rank keep
their physical slots, while incoming experts fill slots vacated by outgoing
experts. Thus, an accepted swap changes only its selected experts; all other
rank--slot assignments remain resident. The resulting placement remains fixed throughout data-layout search
and the associated forward--backward interval.

\begin{figure}[!t]
  \centering
  \includegraphics[width=\columnwidth]
    {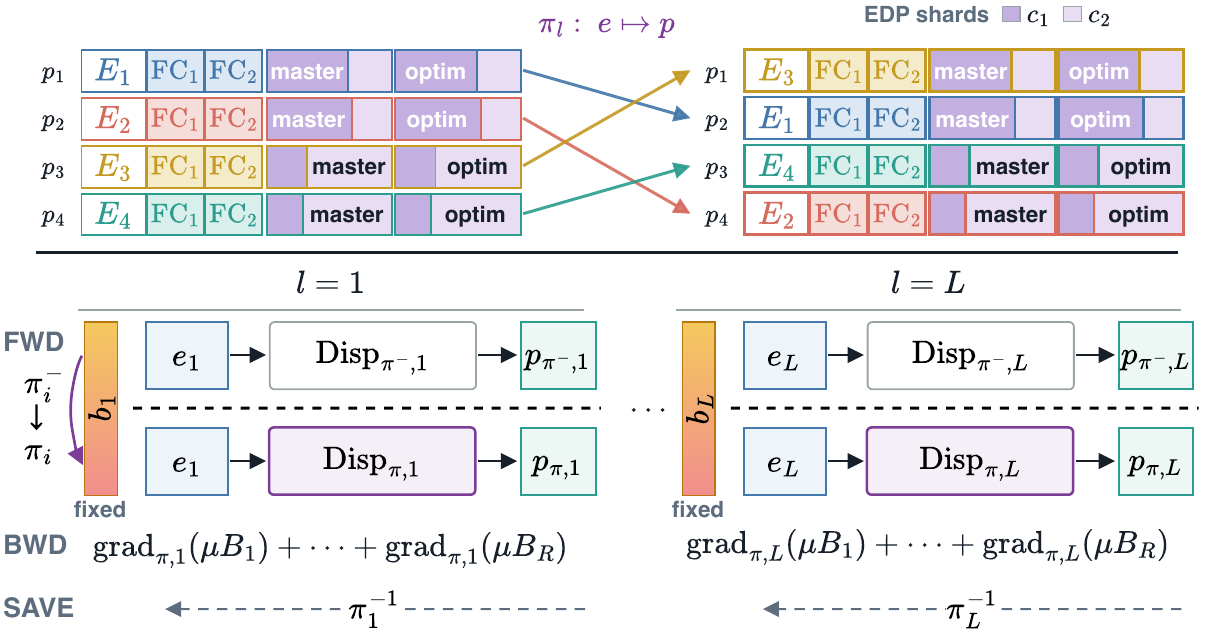}
  \Description{Expert parameters and associated training state are
  materialized under a layer-local permutation. Independent dispatcher maps
  preserve logical routing through forward and backward, while canonical save
  restores logical order.}
  \caption{State-consistent expert placement and canonical export.}
  \label{fig:state-consistency}
\end{figure}

Figure~\ref{fig:state-consistency} summarizes the state-consistent realization
of $\pi_l$. At a training-step boundary, \sysname{} materializes the placement
as part of state loading: expert parameters and their supported expert-indexed
training state are copied into pre-existing physical slots, without replacing
parameter objects or changing tensor shapes, gbuf ranges, or communicator
membership. Logical top-$k$ routing remains unchanged: without changing the router bias,
the dispatcher maps a selected expert $e$ to slot $\pi_l(e)$.
Across all cells $B_{r,d}$ in the $R$-row schedule, the expert mapping stays
fixed during the whole forward--backward interval, allowing backward execution
and activation recomputation to reuse the same physical layout
without rewriting the logical computation graph or its dependencies.
Consequently, 1F1B schedules, gradient accumulation, and the existing parallel
groups retain their original semantics. Appendix~D gives the
detailed state-transfer protocol, incremental placement
pseudocode, and compatibility scope.

\subsection{Controlled Placement Replay}
\label{sec:placement-replay}

We compare the incremental repair above with DeepSeek EPLB$^\ast$ adaptation~\cite{deepseekeplb} driven by the same
aggregated $[L,E]$ demand. The starred R0 row fixes capacity and disables
replicas; R2--R8 add slots and use an idealized even split of each replicated
expert's demand. On the shared replay corpus in Section~\ref{sec:offline-replay-corpus}, each
planner carries its own mapping forward; Table~\ref{tab:planner-comparison}
summarizes 19 transitions after the cold start. All variants are evaluated as offline placement replays under
the same CV target. Timing spans ready loads
to an executable mapping, excluding trace I/O and aggregation.

\begin{table}[!t]
  \caption{Continuous placement replay. Hit:
  layer-steps meeting the CV target; Move: mean logical experts (incremental/R0)
  or replicated physical slots rewritten per transition.}
  \label{tab:planner-comparison}
  \centering
  \RoutePackTableFont
  \begin{tabular*}{\columnwidth}{@{\extracolsep{\fill}}lrrrr@{}}
    \toprule
    Planner & Hit (\%) & Move & Churn (\%) & P50 (ms) \\
    \midrule
    \multicolumn{5}{c}{\textbf{Tiny} ($23\times128$)} \\
    \addlinespace[1pt]
    \sysname{} incremental & 100.00 & 389 & 13.23 & 55.6 \\
    EPLB$^\ast$ R0 & 45.08 & 2,504 & 85.07 & 14.8 \\
    EPLB R2 (+12.5\%) & 53.55 & 2,803 & 84.62 & 18.0 \\
    EPLB R4 (+25.0\%) & 51.95 & 3,028 & 82.27 & 20.4 \\
    EPLB R8 (+50.0\%) & 52.63 & 3,420 & 77.45 & 25.0 \\
    \midrule
    \multicolumn{5}{c}{\textbf{Flash} ($40\times512$)} \\
    \addlinespace[1pt]
    \sysname{} incremental & 100.00 & 736 & 3.59 & 237.0 \\
    EPLB$^\ast$ R0 & 100.00 & 17,877 & 87.29 & 88.5 \\
    EPLB R2 (+3.1\%) & 100.00 & 18,342 & 86.85 & 104.0 \\
    EPLB R4 (+6.3\%) & 100.00 & 18,730 & 86.08 & 109.0 \\
    EPLB R8 (+12.5\%) & 100.00 & 19,509 & 84.67 & 119.0 \\
    \bottomrule
  \end{tabular*}
\end{table}

On Tiny, incremental repair meets the target throughout, versus 45--53\% for
full-rebuild EPLB; it moves $6.4\times$ fewer experts than R0 and retains
86.77\% of the placement. On Flash, all variants meet the target, but incremental
repair moves $24.3\times$ fewer experts than R0 and retains 96.41\%. Replicas
do not improve the hit rate on Flash and add only about eight percentage points on Tiny
while rewriting more slots.

EPLB optimizes worst-rank load through full rebuild rather than our CV gate,
so its lower Tiny hit rate reflects an objective mismatch. Under the common CV
target, EPLB$^\ast$ R0 plans $2.7$--$3.8\times$ faster, but both are sub-second,
and online placement exposes at most 61~ms median wait
(Section~\ref{sec:runtime-costs}). \sysname{}
stops at the CV target and preserves resident expert parameters and optimizer
state; fewer changed experts can reduce state-materialization traffic at
admission~\cite{zhang2025demi}.

\begin{figure*}[!t]
  \centering
  \includegraphics[width=\textwidth]{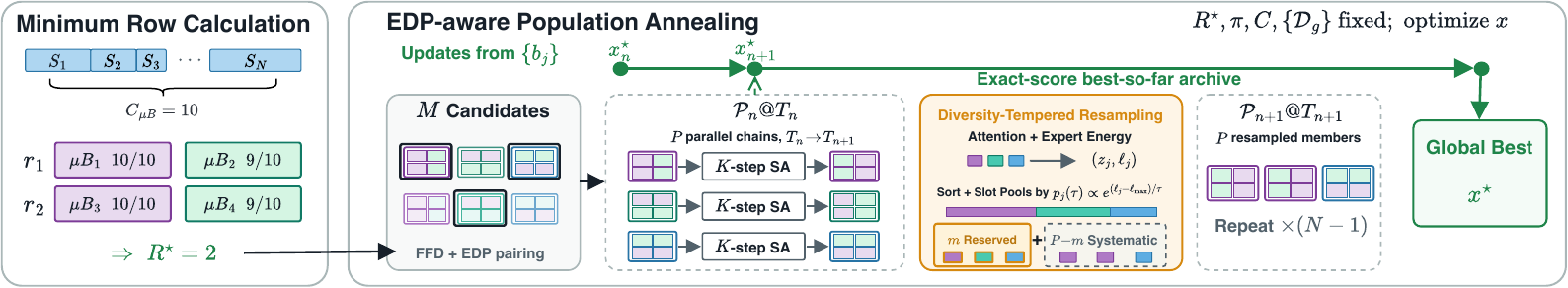}
  \Description{\sysname{} fixes the minimum-certified or best-known row count,
  constructs a diverse feasible population, and executes parallel simulated-
  annealing chains. Protected top-m slots and systematic residual slots form
  the next population, while per-chain best states update a separate
  monotonic archive.}
  \caption{Fixed-row rescheduling: diverse seeding, parallel annealing, and
  protected-parent resampling with a best-so-far archive.}
  \label{fig:joint-planning}
\end{figure*}

\subsection{Stage-Aware Objective and Scope}
\label{sec:formulation-guarantees}

The normalized cell statistics are

\begin{RoutePackCompactDisplay}
\begin{equation}
  \bar T_{r,d}=\frac{1}{C}\sum_i t_i x_{i,r,d},\qquad
  \bar Q_{r,d}=\sum_i(\frac{t_i}{C})^2x_{i,r,d}.
  \label{eq:cell-token-statistics}
\end{equation}
\end{RoutePackCompactDisplay}

For each dense-side stage $s$ (attention or dense FFN), its token-linear and token-pair terms are
combined per DP cell before taking the slowest cell in EDP shard $g$. The
separate stage maxima are then accumulated because the stages lie serially on
the projected row critical path:

\begin{RoutePackCompactDisplay}
\begin{equation}
  A_{r,g}=\sum_{s\in\mathcal A}
  \max_{d\in\mathcal D_g}
  (\alpha_s\bar T_{r,d}+\beta_s\bar Q_{r,d}).
  \label{eq:stage-aware-attention}
\end{equation}
\end{RoutePackCompactDisplay}

The corresponding row-local EP load is

\begin{RoutePackCompactDisplay}
\begin{equation}
  W_{r,g,l,p}(\pi,x)=
  \sum_{d\in\mathcal D_g}\sum_i x_{i,r,d}q_{i,l,p}(\pi_l),
  \label{eq:row-expert-load}
\end{equation}
\end{RoutePackCompactDisplay}

Summing the busiest physical rank independently at every MoE layer gives

\begin{RoutePackCompactDisplay}
\begin{equation}
  E_{r,g}=\frac{1}{C}\sum_{l\in\mathcal L_{\mathrm{MoE}}}
  \max_p W_{r,g,l,p}(\pi,x).
  \label{eq:row-expert-cost}
\end{equation}
\end{RoutePackCompactDisplay}

Let $J_{r,g}=A_{r,g}+E_{r,g}$ and $U_g=\sum_rJ_{r,g}$. At fixed
$(R^\star,\pi)$, \sysname{} ranks layouts by

\begin{RoutePackCompactDisplay}
\begin{equation}
  \operatorname{Score}(x;\pi)=
  \left(\max_gU_g,\;\sum_gU_g,\;\max_{r,g}J_{r,g}\right),
  \label{eq:lexicographic-score}
\end{equation}
\end{RoutePackCompactDisplay}

ordered lexicographically. The primary term controls the slowest EDP shard
over the optimizer step; total projected work and the worst row-local tail
break ties. Conceptually, the planner minimizes $(R,\operatorname{Score})$;
operationally, it fixes $R^\star$ before comparing layouts.

For $G$ equal-sized EDP shards, an assignment-independent fractional lower
bound on the primary cost $J=\max_g U_g$ is
\begin{RoutePackDenseDisplay}
\begin{equation}
\begin{aligned}
  B_{\mathrm{EP}}&=\frac{1}{GC}
    \sum_{l\in\mathcal L_{\mathrm{MoE}}}\max_p\sum_i q_{i,l,p},\\
  B_{\mathrm{dense}}&=\frac{1}{D}\sum_{s\in\mathcal A}
    \left[\frac{\alpha_s}{C}\sum_i t_i
    +\beta_s\sum_i(t_i/C)^2\right],\\
  B&=B_{\mathrm{EP}}+B_{\mathrm{dense}}.
\end{aligned}
  \label{eq:fractional-rescheduling-bound}
\end{equation}
\end{RoutePackDenseDisplay}
It permits fractional splitting across cells and shards, so $J/B\geq1$;
integer capacity and whole-trajectory assignment can make equality unattainable.

Every construction and local move in this comparison preserves
Equations~\ref{eq:trajectory-bijection}--\ref{eq:token-capacity}, the selected row
count, and the EDP topology. Exact lexicographic comparison maintains the best
feasible incumbent.

Planning begins only after cross-trajectory statistics required by the learning
rule have been attached to their records. Reordering then preserves every
record and the complete optimizer-step multiset. It leaves the mathematical
update unchanged whenever that update is invariant to record ordering and
physical ownership; order-dependent algorithms must materialize their
dependencies before planning.

\newcommand{\opWindowShuffle}{\textsc{Window\allowbreak Shuffle}\xspace}
\newcommand{\opRandomizedBestFit}{\textsc{Randomized\allowbreak Best\allowbreak Fit}\xspace}
\newcommand{\opEDPAwarePair}{\textsc{EDP\allowbreak Aware\allowbreak Pair}\xspace}
\newcommand{\opCanonicalize}{\textsc{Canonicalize}\xspace}
\newcommand{\opCanonicalMap}{\textsc{Canonical\allowbreak Map}\xspace}
\newcommand{\opQualityDiversitySelect}{\textsc{Quality\allowbreak Diversity\allowbreak Select}\xspace}
\newcommand{\opGuidedProposal}{\textsc{Guided\allowbreak Proposal}\xspace}
\newcommand{\opIncrementalScore}{\textsc{Incremental\allowbreak Score}\xspace}
\newcommand{\opNormalize}{\textsc{Normalize}\xspace}

\section{Joint Dense-Side and Expert Trajectory Rescheduling}
\label{sec:solver}

After fixing $(R^\star,\pi)$, \sysname{} jointly optimizes the dense-side and
expert components of Equation~\ref{eq:lexicographic-score} by searching only
trajectory-to-cell assignments in $\mathcal X_{R^\star}$. Figure~\ref{fig:joint-planning}
and Algorithm~\ref{alg:routepack-planner} summarize the production search,
which combines feasible construction, diversity-aware population selection,
and parallel population annealing.

\begin{figure*}[!t]
  \centering
  \includegraphics[width=\textwidth]{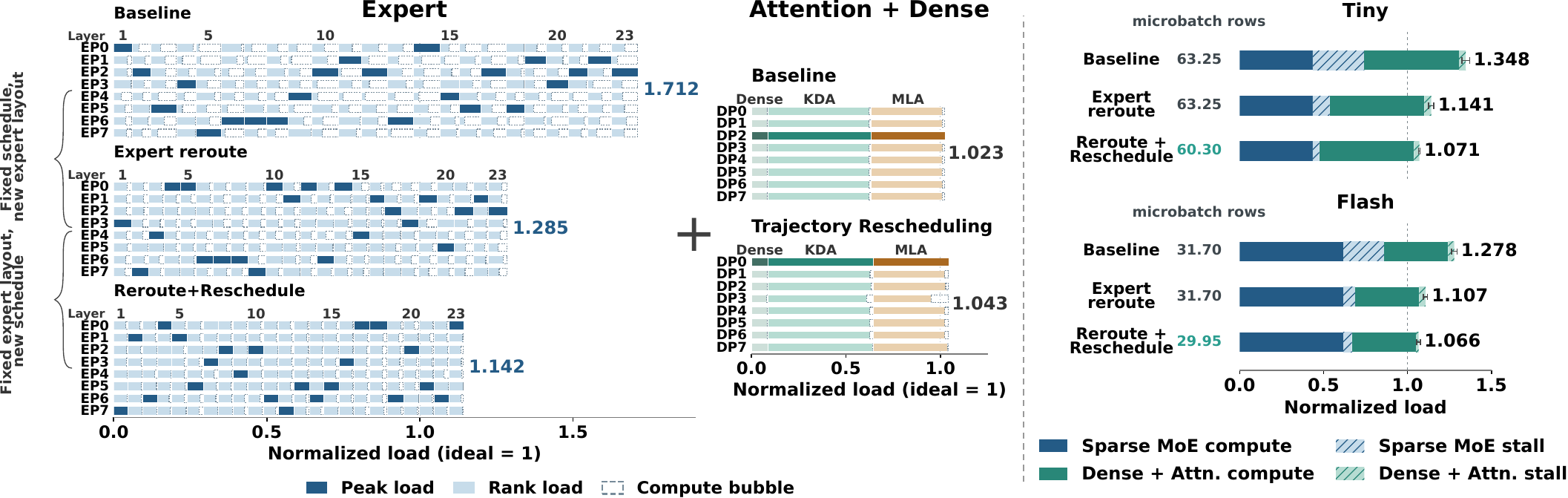}
  \Description{Left: Three expert-load profiles compare baseline, expert
  rerouting, and rerouting with trajectory rescheduling; two dense-side profiles compare
  the baseline schedule with the rescheduled Tiny trajectory layout. Right: Three bars per model
  summarize sparse and dense-side compute, peak-minus-mean gaps, and
  method-aligned mean row counts over 20 windows.}
  \caption{Modeled computation workloads. Left: one randomly selected non-tail
  Tiny execution row on separate expert and dense-side row-mean scales. Right:
  20-window Tiny/Flash means; solid segments are mean compute, hatching is the
  peak-minus-mean gap, and whiskers show standard deviation.}
  \label{fig:load-breakdown}
\end{figure*}

\begin{figure}[!t]
  \centering
  \includegraphics[
    width=\columnwidth,
  ]{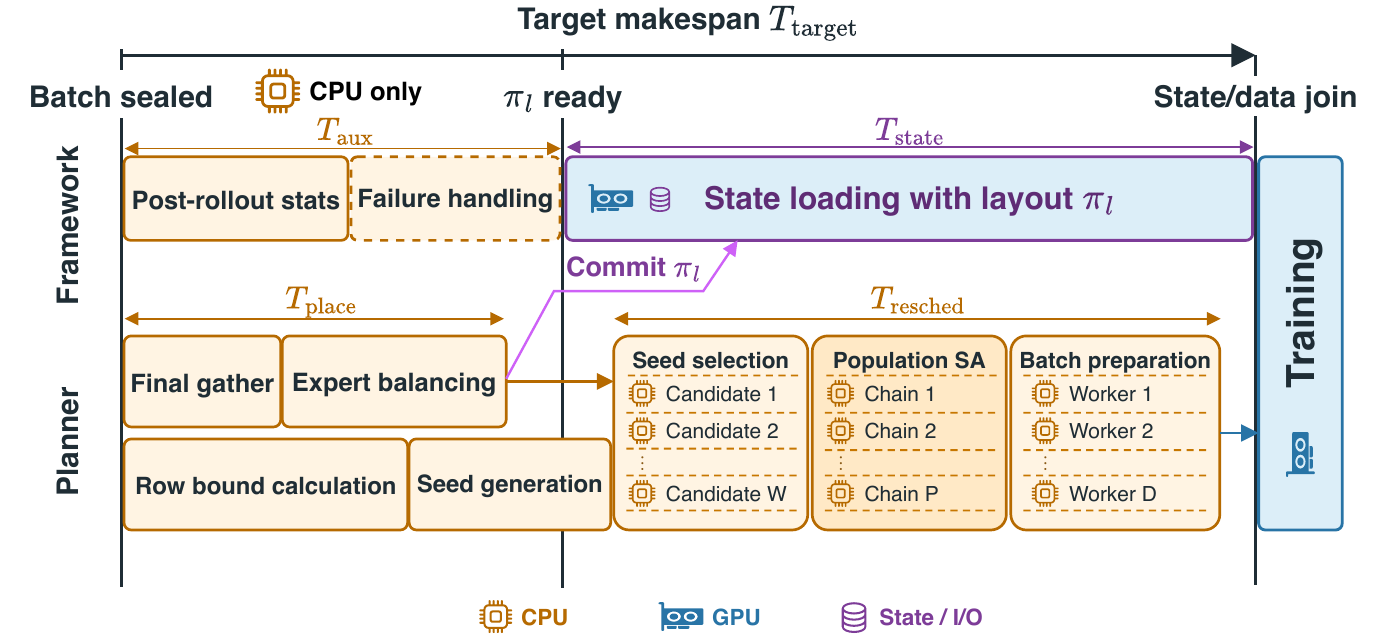}
  \Description{A target two-lane schedule of framework execution and CPU-side
  planning. Post-rollout work, optional failure handling, and
  placement-aware state loading form the framework path. Routing aggregation,
  expert balancing, seed construction,
  population annealing, and batch preparation form the planner path. State and
  prepared data join before GPU training.}
  \caption{Target post-rollout overlap schedule at training admission.}
  \label{fig:runtime-overlap}
\end{figure}

\subsection{Population Construction and Parallel Annealing}

\paragraph{Feasible population.}
Length-FFD~\cite{johnson1974binpacking} first establishes a feasible layout
with $R^\star D$ cells. \sysname{}
retains it and an EDP-aware pairing of the same cells, then perturbs decreasing
length order and chooses among the most occupied feasible cells. It pairs
completed cells into rows by their incremental layer-wise EDP-rank peaks and
rejects any trial that violates capacity, empties a cell, or changes trajectory
coverage.

Canonicalization removes symmetries from row order, trajectory order within a
cell, and exchangeable DP slots inside an EDP shard. Population selection keeps
the exact-score winner, then adds competitive, structurally distinct candidates
using co-assignment and compact token/load projections. Appendix~E
gives the complete construction and distance definition.

\paragraph{Parallel annealing.}
Each population member runs an independent annealing chain over a geometric
temperature segment~\cite{kirkpatrick1983sa,hukushima2003population}. A proposal
either swaps trajectories between cells or relocates one trajectory
without emptying its source. Most proposals choose a high-cost source
row--shard and a low-cost target; the remainder preserve random exploration.
Incremental rescoring updates only the two affected cells and their associated
row--shard expert loads.

At a segment boundary, path weights favor chains that remain strong throughout
cooling. Tempering limits population collapse, protected parents retain the
best distinct terminal states, and systematic resampling~\cite{carpenter1999resampling} fills the remaining
workers. A separate global incumbent is updated from every per-chain best
state using the exact score in Equation~\ref{eq:lexicographic-score}. The
returned layout is validated and cannot score worse than the best initial
seed. Population members execute in a persistent CPU process pool, so one
segment is paced by its slowest chain rather than by the sum of all proposals.

\begin{routealgorithmfloat}{\sysname{} hierarchical planning}
  \label{alg:routepack-planner}
  \AlgRequire{Replay profiles $\{(t_i,\mathbf A_i)\}$; capacity $C$; DP and EDP
  topology; search budget}
  \AlgEnsure{Layer maps $\pi$ and feasible layout $x^\star$}
  \AlgLine{Aggregate $L_{l,e}\gets\sum_i a_{i,l,e}$ and incrementally repair
    each carried placement $\pi_l^{-}$ to obtain $\pi_l$}
  \AlgLine{Find the minimum-certified or best-known feasible $R^\star$}
  \AlgLine{Construct canonical fixed-row candidates from FFD, randomized
    best-fit, and EDP-aware cell pairing}
  \AlgLine{$\mathcal P_0\gets\opQualityDiversitySelect(\mathcal C)$;
    $x^\star\gets\arg\min_{x\in\mathcal P_0}\operatorname{Score}(x;\pi)$}
  \AlgLine{Run parallel guided swap/relocation chains over each temperature
    segment; update $x^\star$ from all per-chain best states}
  \AlgLine{Temper and resample terminal states while protecting strong
    distinct parents}
  \AlgLine{Validate $x^\star\in\mathcal X_{R^\star}$ and return $(\pi,x^\star)$}
\end{routealgorithmfloat}

Appendix~E specifies geometric cooling, path weighting,
resampling, and the complete pseudocode.

\subsection{Controlled Replay and Framework Integration}
\label{sec:mechanism-decomposition}

We replay both controls on the shared routing corpus to verify and visualize
their effects on modeled dense-side and expert workloads before online measurement.

Figure~\ref{fig:load-breakdown} compares the same three policies as the online
evaluation in Section~\ref{sec:evaluation} (DP=EP=8, EDP=1; 8192-token cap).
\textbf{Baseline} (\textbf{Base} online) keeps identity placement and uses
DP-local FFD: assign consecutive eight-trajectory shards to DP slots, shuffle
equal-length ties, and pack decreasing lengths into the feasible cell with the
fewest trajectories, opening a new cell if needed; row counts are synchronized.
Expert reroute (\textbf{Reroute}) remaps each layer's experts to physical
EP ranks, leaving logical routing unchanged, while retaining Baseline's packed
rows; \textbf{Full} (Reroute+Reschedule) also reschedules the complete window
at the window-level row count $R^\star$.

Section~\ref{sec:background-insufficient} predicts the control order:
rescheduling cannot lower the placement-defined global EP bound, whereas
rerouting can. To test both effects, we also replay \textbf{Reschedule}, Full's
data rescheduling under identity placement. Uniform EP work is

\begin{RoutePackCompactDisplay}
\begin{equation*}
  U=\frac{1}{8}\sum_l\sum_p G_{l,p}.
\end{equation*}
\end{RoutePackCompactDisplay}

Table~\ref{tab:global-local-replay} separates the two EP excess terms:

\begin{RoutePackCompactDisplay}
\begin{equation*}
  \begin{aligned}
    \text{Global}   &= \frac{\sum_l\max_p G_{l,p}-U}{U},\\
    \text{Row-local}&= \frac{\sum_{r,l}\max_p W_{r,l,p}-\sum_l\max_p G_{l,p}}{U}.
  \end{aligned}
\end{equation*}
\end{RoutePackCompactDisplay}

As predicted, Reschedule leaves global excess unchanged; reducing only the
row-local term yields 4.18\% and 2.50\% joint modeled savings on Tiny and Flash.
Reroute nearly removes global excess, exposing a larger row-local gap for Full
to reduce; Full's joint savings reach 20.56\% and 16.61\%, respectively.
Their modeled savings are nearly additive, consistent with distinct global
and row-local targets.

\begin{table}[!t]
  \caption{Global and row-local EP excess; joint cost saved versus Base
  (20-window means, \%).}
  \label{tab:global-local-replay}
  \centering
  \RoutePackTableFont
  \setlength{\tabcolsep}{3pt}
  \begin{tabular*}{\columnwidth}{@{\extracolsep{\fill}}llrrr@{}}
    \toprule
    Model & Policy & Global & Row-local & Joint saved \\
    \midrule
    Tiny  & Base    & 61.23 &  9.40 &  0.00 \\
          & Reschedule    & 61.23 &  1.41 &  4.18 \\
          & Reroute &  0.06 & 23.17 & 15.30 \\
          & Full    &  0.06 &  8.84 & 20.56 \\
    \addlinespace[2pt]
    Flash & Base    & 36.03 &  3.18 &  0.00 \\
          & Reschedule    & 36.03 &  2.23 &  2.50 \\
          & Reroute &  0.05 & 11.39 & 13.41 \\
          & Full    &  0.05 &  8.24 & 16.61 \\
    \bottomrule
  \end{tabular*}
\end{table}

The selected non-tail Tiny row trades a small dense-side peak increase for a
larger expert-peak reduction. Across 40 windows, Full lowers the joint modeled
cost and both gaps; Tiny's residual sparse and dense-side gaps become
comparable, while Flash remains sparse-dominated. Row compaction alone does
not produce these modeled savings: under the same placement, global
length-only FFD at $R^\star$ raises the joint cost by 3.35\% (Tiny) and
1.04\% (Flash) relative to DP-local FFD, whereas \sysname{}'s refinement at
that row count lowers it by 9.42\% and 4.68\% relative to global FFD
(Appendix~K).

Rollout replay captures routed experts as trajectories complete.
Figure~\ref{fig:runtime-overlap} sketches the target: post-seal profile
finalization and placement repair overlap post-rollout work; after $\pi$ is
ready, CPU rescheduling and versioned state preparation run in parallel and
join at training admission.

Let $T_{\mathrm{place}}$ cover post-seal profile finalization and placement repair,
$T_{\mathrm{aux}}$ the post-rollout statistics and safety gates, and
$T_{\mathrm{state}}$ placement-aware state loading. The data branch
$T_{\mathrm{resched}}$ comprises asynchronous CPU planning
($T_{\mathrm{plan}}$) and subsequent worker batch preparation. The
modeled sealing-to-admission makespan for this target schedule is

\begin{RoutePackCompactDisplay}
\begin{equation}
  T_{\mathrm{target}}=\max\{T_{\mathrm{place}}+T_{\mathrm{resched}},
  \max(T_{\mathrm{place}},T_{\mathrm{aux}})+T_{\mathrm{state}}\}.
  \label{eq:target-runtime-dag}
\end{equation}
\end{RoutePackCompactDisplay}

The rescheduling branch therefore stays off the modeled admission path whenever
the first term does not exceed the second.

Appendix~F gives host complexity and worker bounds.

\subsection{Comparison with ILP Solver}
\label{sec:ilp-comparison}

Section~\ref{sec:fixed-row} formulates fixed-row trajectory scheduling as a
0--1 ILP, making its optimum exactly solvable in principle. We
solve it offline with Google OR-Tools CP-SAT~\cite{ortools} as a generic
diagnostic. CP-SAT presolves the integer model, then runs a parallel portfolio
with conflict learning, propagation, linear relaxations, and neighborhood
search. Unlike Population-SA, it can return
a certified lower bound or prove optimality, but its assignment space grows as
$O(NRD)$ before the auxiliary peak variables are counted.

The comparison fixes placement, row count, capacity, and topology, and
minimizes only $\max_gU_g$ from Equation~\ref{eq:lexicographic-score}. The
common global length-only (FFD) layout supplies a complete assignment hint
and a safe objective cut. The hint guides search but is not itself a registered CP-SAT
incumbent; consequently, an \textsc{Unknown} status means that the solver
returned no feasible assignment within the budget, not that the fixed-row
problem is infeasible.

Table~\ref{tab:ilp-scaling} compares reduction from that FFD layout on five warm
profiles per model with complete eight-trajectory GRPO groups. The nested
prefixes use $N\in\{32,64,128,256\}$. Population-SA has 64 members, two
temperature levels, and 10,000 proposals per member and level; both methods use
at most 64 workers. CP-SAT has a 60-s limit including presolve, longer than
the 8K online median $T_{\mathrm{state}}$ (Tiny: 7.39~s; Flash: 34.21~s;
Table~\ref{tab:planning-state-overlap}). Appendix~J reports fixed-point
quantization, statuses, and certified bounds.

\begin{table}[!t]
  \centering
  \caption{Median fixed-row primary-objective reduction from the common FFD
  layout over five profiles (\%; higher is better). A dash means no incumbent within 60 seconds.}
  \label{tab:ilp-scaling}
  \RoutePackTableFont
  \setlength{\tabcolsep}{2.5pt}
  \begin{tabular*}{\columnwidth}{@{\extracolsep{\fill}}rrrrr@{}}
    \toprule
    $N$ & Tiny SA & Tiny ILP & Flash SA & Flash ILP \\
    \midrule
     32 &  6.63 & 6.64 & 12.02 & 11.59 \\
     64 &  5.68 & 5.35 &  8.38 &  6.77 \\
    128 &  7.53 & 1.04 &  6.73 &  2.38 \\
    256 &  6.19 &   -- &  4.99 &    -- \\
    \bottomrule
  \end{tabular*}
\end{table}

At $N=32$, only one Tiny profile is certified optimal; none of the five Flash
profiles is. No profile is certified optimal at larger $N$. By $N=128$, CP-SAT's
median reduction is 1.04--2.38\%, versus
6.73--7.53\% for Population-SA. At $N=256$, CP-SAT registers no incumbent in
any selected profile, whereas Population-SA still improves FFD by 4.99--6.19\%
and retains a valid incumbent throughout its budget. Thus the generic solver
is useful as a small-instance reference, but becomes uneconomical when the production window scales up.

The sweep stops at $N=256$: a 512-trajectory solve or longer budget could
improve or certify an incumbent, but would exceed the bounded online search
budget.

\section{Evaluation}
\label{sec:evaluation}

This section reports online Megatron throughput in six configurations, plan
quality, and admission timing in real post-training RL tasks.

\subsection{Setup and Measurement Protocol}

We evaluate the Tiny and Flash models from
Section~\ref{sec:offline-replay-corpus} in an in-house RL
stack. All three policies use the same batching, state-management, routing
replay, and Megatron execution interfaces.\footnote{We omit direct online
comparisons with ReLibra and ForeMoE~\cite{relibra,foremoe}: public
implementations were unavailable during evaluation, and their microbatch-level
replication and token assignment require different execution and
state-management support.} Every window holds 512
trajectories as in Section~\ref{sec:offline-replay-corpus}. Reroute and Full additionally
aggregate the replayed expert IDs into one $[L,E]$ routing matrix per
trajectory as rollouts complete.

Table~\ref{tab:evaluation-setup} lists shared controls and per-run parallelism
and rollout settings; CP and expert TP are one throughout. EDP
counts planner-visible shards in the runtime partition $\{\mathcal D_g\}$, not
additional GPUs. H20-3e uses all rows, while the cross-GPU study reuses the Tiny
16K row. Calibration is fixed for each configuration and hardware;
Appendix~K reports its coefficients and normalization.

\begin{table}[!t]
  \centering
  \caption{Runtime, parallelism, and rollout settings by model and context cap.}
  \label{tab:evaluation-setup}
  \RoutePackTableFont
  {\setlength{\fboxsep}{3pt}%
  \colorbox{black!4}{%
    \begin{tabular}{@{}ll@{}}
      \texttt{precision:} & \texttt{BF16} \\
      \texttt{moe\_kernel:} & \texttt{grouped GEMM} \\
      \texttt{dispatcher:} & \texttt{DeepEP flex} \\
      \texttt{routing\_groups:} & \texttt{8} \\
      \texttt{group\_topk:} & \texttt{4} \\
      \texttt{expert\_topk:} & \texttt{8} \\
      \texttt{updates\_per\_window:} & \texttt{1} \\
      \texttt{attention\_backend:} & \texttt{FlashAttention-3}~\cite{shah2024flashattention3}  \\
      \texttt{training\_runtime:} & \texttt{Megatron-LM}~\cite{narayanan2021megatron} \\
      \texttt{recompute:} & \texttt{Attention, MoE/MLP activations} \\
      \texttt{loss\_normalization:} & \texttt{global}\\
    \end{tabular}%
  }}
  \vspace{2pt}

  \setlength{\tabcolsep}{2.5pt}
  \begin{tabular*}{\columnwidth}{@{\extracolsep{\fill}}lrrrrrr@{}}
    \toprule
    Run & DP & TP & PP & EP & EDP & Cap. \\
    \midrule
    Flash 8K  & 8 & 1 & 4 & 8 & 1 & 8,192 \\
    Flash 16K & 8 & 2 & 4 & 8 & 2 & 16,383 \\
    Flash 32K & 8 & 2 & 4 & 8 & 2 & 32,767 \\
    Tiny 8K   & 8 & 1 & 1 & 8 & 1 & 8,192 \\
    Tiny 16K  & 8 & 1 & 1 & 8 & 1 & 16,384 \\
    Tiny 32K  & 8 & 2 & 1 & 8 & 1 & 32,768 \\
    \bottomrule
  \end{tabular*}
\end{table}

Here \texttt{updates\_per\_window: 1} means that each sealed optimizer window
receives exactly one gradient update rather than PPO-style multiple updates. We normalize the loss over the
full optimizer step, so the loss does not depend on the trajectory schedule.

\textbf{Base}, \textbf{Reroute}, and \textbf{Full} apply the policies defined in
Section~\ref{sec:mechanism-decomposition} to live sealed windows; unlike the
offline reconstruction, Base uses its actual shuffle state. Reroute keeps
Base's DP-local rows, whereas Full uses the window-level row count $R^\star$. We collect
online throughput during separate continuous mixed-task runs under the same deployment
configuration, covering the post-admission Megatron training region rather
than an isolated MoE operator. Following ReLibra~\cite{relibra}, we report
training-stage throughput because \sysname{}'s plan changes how the training
stage executes; its admission-path costs are reported separately in
Section~\ref{sec:runtime-costs}.

For the mechanism analysis in Section~\ref{sec:trace-attribution}, we
additionally collect a separate all-rank Tiny 16K device trace for each method.

\subsection{Online Megatron Throughput}
\label{sec:training-throughput}

\subsubsection{Across Different Context Windows}
\label{sec:window-scaling}

For each deployment, we compare a common chronological prefix of the three
jobs: logical step~1 is warmup, followed by 99 updates (98 for Tiny 16K).
Repeated logger labels are ordered chronologically. We report
$\sum_i T_i/\sum_i \tau_i$, where $T_i$ is the effective training-token count and
$\tau_i$ is Megatron training-region time, thereby weighting each processed token
equally rather than averaging rates from unequal updates.

\begin{figure}[t]
  \centering
  \includegraphics[width=\columnwidth]{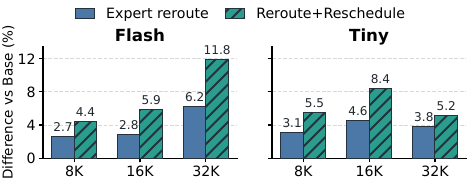}
  \Description{Two grouped bar charts show throughput differences from the baseline
  for expert rerouting and rerouting plus trajectory rescheduling on Flash and Tiny at 8K,
  16K, and 32K context caps.}
  \caption{Observed Megatron training-region TPS difference from Base across
  six H20-3e configurations.}
  \label{fig:training-throughput}
\end{figure}

Figure~\ref{fig:training-throughput} shows the same ordering in all six
deployment runs: expert placement alone (Reroute) raises TPS by 2.67--6.24\%
over Base, and adding trajectory rescheduling (Full) raises it by
4.44--11.85\% over Base and by 1.33--5.28\% over Reroute. The ordering also
holds at intermediate 10-, 20-, 40-, 60-, and 80-update prefixes.
The Full--Reroute contrast measures deployed training schedules, including
changes in compute, collectives, and synchronization; it combines row
compaction to $R^\star$ with routing-aware refinement. At fixed $R^\star$,
refinement alone lowers the modeled joint cost of the online plans by
2.11--5.06\% (Section~\ref{sec:runtime-costs}).
Across the same configurations, both controls show reward progression
comparable to Base over the first 100 optimizer updates (Appendix~G).

\subsubsection{Across Different GPUs}

We additionally post-train the Tiny model with 16K context on NVIDIA
Hopper H20 and H200 GPUs. Each GPU exposes 900~GB/s nominal bidirectional
NVLink~\cite{nvidiahopper} but differs in dense BF16 throughput and HBM
capacity. Training configurations remain the same as in
Section~\ref{sec:window-scaling}.

Table~\ref{tab:accelerator-check} reports raw token-weighted TPS. Full improves
same-platform training-region TPS by 1.92--8.37\% across all three GPUs.
Section~\ref{sec:communication-pressure} analyzes this cross-platform trend.

\begin{table}[t]
  \centering
  \caption{Tiny 16K cross-GPU check. TPS is in
  kTok/s; $\Delta_{\mathrm{Full}}$ uses the same-GPU Base. Parentheses report
  dense BF16 TFLOPS and HBM capacity (GB).}
  \label{tab:accelerator-check}
  \RoutePackTableFont
  \setlength{\tabcolsep}{3pt}
  \begin{tabular*}{\columnwidth}{@{\extracolsep{\fill}}lrrrr@{}}
    \toprule
    GPU & Base & Reroute & Full & $\Delta_{\mathrm{Full}}$ \\
    \midrule
    H20-3e (148/141) & 47.119 & 49.286 & 51.064 & +8.37\% \\
    H200 (989/141)   & 92.401 & 92.109 & 94.177 & +1.92\% \\
    H20 (148/96)     & 42.396 & 43.459 & 44.088 & +3.99\% \\
    \bottomrule
  \end{tabular*}
\end{table}

\subsection{Planner Quality and Admission Timing}
\label{sec:runtime-costs}

\paragraph{Plan quality.}
For each online input, we compare the final plan with its global length-only
(FFD) initialization at fixed placement, trajectories, and row count,
isolating routing-aware refinement after compaction. Let $J_0$ and $J_1$ be the initial
and final primary costs, normalized by $B$ from
Equation~\ref{eq:fractional-rescheduling-bound}. Over 100 plans per
configuration, we average $J_0/B$,
$J_1/B$, excess removal $\eta=(J_0-J_1)/(J_0-B)$, and reduction
$1-J_1/J_0$ per plan. Here $\eta$ measures search quality, not recoverable
execution time.

\begin{figure}[t]
  \centering
  \includegraphics[width=\columnwidth]{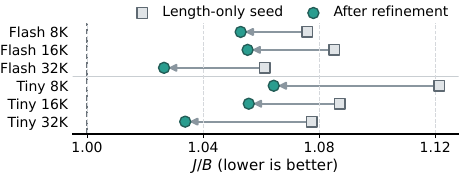}
  \Description{Six dumbbell rows show the joint cost of online plans relative
  to their fractional lower bound before and after routing-aware trajectory
  rescheduling. Every refined point is closer to one.}
  \caption{Online plan refinement of the modeled objective. Arrows point toward
  the fractional lower bound.}
  \label{fig:online-packing-quality}
\end{figure}

Figure~\ref{fig:online-packing-quality} shows that refinement removes
28.9--54.3\% of the initial excess across all recorded online plans
and lowers $J$ by 2.11--5.06\%. Across configurations, the expert component falls by
2.6--24.8\%, while dense-side cost changes by $-0.4$\% to $+5.4$\%; the joint search
can therefore accept a small dense-side increase to remove a larger sparse tail.
For EDP=2, $J$ uses the maximum shard total rather than summing shard-local
component reports.

\paragraph{Planning and admission timing.}
Against Figure~\ref{fig:runtime-overlap}'s target schedule, post-seal profile
finalization plus placement repair takes 0.10--0.39~s p50 and finishes within
$T_{\mathrm{aux}}$ in 389/594 warm windows (65.5\%). Its exposed wait beyond
$T_{\mathrm{aux}}$ (gate tail) has a p50 of zero in four runs and 50--61~ms in
the other two; seven Flash windows have positive tails despite p95 zero. Commit takes
0.23--3.12~s p50.

CPU rescheduling takes 2.18--3.25~s p50, with 97.4--99.4\% median
planner/state overlap. State loading takes 4.91--34.21~s p50. Post-state
batch await takes 27--55~ms on Flash versus 2.31--4.16~s on Tiny, including
worker readiness and signaling during batch preparation. Base is the
unmodified production admission path, which prepares batches and loads state
without overlapping them, so it has no counterpart to the planner/state
overlap measured here.

\begin{table}[H]
  \centering
  \caption{Warm-window medians (s). Gate tail follows
  $T_{\mathrm{aux}}$; batch await follows state loading.}
  \label{tab:planning-state-overlap}
  \RoutePackTableFont
  \setlength{\tabcolsep}{2pt}
  \begin{tabular*}{\columnwidth}{@{\extracolsep{\fill}}lrrrrr@{}}
    \toprule
    Run & $T_{\mathrm{aux}}$ & Gate tail & $T_{\mathrm{plan}}$ & $T_{\mathrm{state}}$ & Batch await \\
    \midrule
    Flash 8K  & 1.82 & 0 & 2.18 & 34.21 & 0.027 \\
    Flash 16K & 1.01 & 0 & 2.27 & 31.65 & 0.055 \\
    Flash 32K & 1.01 & 0 & 2.67 & 31.21 & 0.055 \\
    Tiny 8K   & 0.045 & 0.061 & 3.25 & 7.39 & 2.418 \\
    Tiny 16K  & 2.02 & 0 & 2.84 &  9.08 & 4.165 \\
    Tiny 32K  & 0.052 & 0.050 & 2.33 &  4.91 & 2.309 \\
    \bottomrule
  \end{tabular*}
\end{table}

\section{Discussion and Limitations}
\label{sec:discussion}

\subsection{Modeling EP Communication Overhead}
\label{sec:communication-pressure}

Standard expert parallelism dispatches routed activations to expert-owning
ranks and combines the outputs through a second all-to-allv
exchange~\cite{lepikhin2021gshard,jin2026megascale}. For each execution row
and MoE layer in EDP shard $g$, we construct the off-rank traffic matrix
$M_{r,g,l}$ induced by the recorded routing demand, the row assignment, and
the physical expert placement. Let $V_{r,l}$ be its total off-rank demand and
$m_{r,l}$ its largest send-or-receive endpoint load. SABRE's endpoint-capacity
argument~\cite{wei2026sabre} motivates comparing $m_{r,l}$ with the balanced
endpoint load $V_{r,l}/n$ for an EP group of $n$ ranks. We aggregate this
diagnostic across rows and layers as
\begin{RoutePackCompactDisplay}
\begin{equation}
  \rho=\frac{n\sum_{r,l}m_{r,l}}{\sum_{r,l}V_{r,l}},
  \label{eq:dispatch-pressure}
\end{equation}
\end{RoutePackCompactDisplay}
Thus $\rho - 1$ describes the excess endpoint pressure. Appendix~B gives the full dispatch--combine derivation and its
service-time assumptions.

We replay the three layouts in Table~\ref{tab:dispatch-demand} on the shared
20-window corpus from Section~\ref{sec:offline-replay-corpus}. All use
DP=EP=8 and EDP=1. The
assignment-level diagnostic retains the
baseline row count and reruns rescheduling at that fixed count, isolating
redistribution from changes in collective count.
For these EDP=1 runs, $n=8$ and each source DP slot is its own EP rank;
Table~\ref{tab:dispatch-demand} reports
$\rho=8\overline{\Psi}_{\mathrm{comm}}/\overline V$, where
$\overline{\Psi}_{\mathrm{comm}}$ and $\overline V$ are the corpus-window means
of the numerator and denominator sums in Equation~\ref{eq:dispatch-pressure}. Mean $V$ changes by less than 0.03\% relative to Baseline.
Expert rerouting removes 62.8\% and 65.9\% of
the modeled imbalance overhead on Tiny and Flash, respectively; joint rescheduling
raises these reductions to 82.6\% and 76.5\%. The improvement comes from lower
endpoint concentration rather than lower assignment volume.

We additionally replay the same corpus with DeepEP's rank-level deduplication:
a token is sent once per destination rank even when several selected experts
are colocated. Under the baseline's group-top-$k=4$ and group-to-rank alignment,
selected groups constrain destination-rank fanout. Rerouting disperses these
experts across ranks and raises remote rank-token copies by 38.3\% on Tiny and
23.3\% on Flash while removing persistent endpoint skew. At fixed placement,
rescheduling preserves window-total destination-rank copies, including local
copies, but can change which copies are off-rank. It redistributes demand
across rows, reducing the
cumulative per-row/layer endpoint peak by 6.7\% and 2.5\% relative to
rerouting-only, leaving it 1.3\% and 4.9\% above Baseline, while raising
peak-weighted endpoint efficiency from 76.7\% and 85.1\% to 93.0\% and 94.9\%.

This locality--balance trade-off helps interpret the cross-GPU trend in
Table~\ref{tab:accelerator-check}. On H200, Reroute is slightly below Base,
whereas Full is above it. Faster compute may expose reroute-induced fanout more
strongly; Full's trajectory schedule can affect compute and collectives through
row count and row-local loads. The H20 and H20-3e results are consistent with
more compute-tail headroom relative to communication pressure. DeepEP also
streams routed activations and temporary buffers through HBM, so equal
nominal NVLink bandwidth does not imply identical
realized transport. H20-3e's larger 141-GB HBM capacity can ease buffer pressure
relative to H20's 96 GB, providing a second plausible contributor to its larger
gain, although these measurements do not isolate capacity from realized HBM
service.

\begin{table}[t]
  \centering
  \caption{Token assignment imbalance and DeepEP rank-deduplicated replay. Changes are relative to Baseline; Eff. is peak-weighted endpoint efficiency.}
  \label{tab:dispatch-demand}
  \RoutePackTableFont
  \setlength{\tabcolsep}{2.5pt}
  \begin{tabular}{@{}llrrrr@{}}
    \toprule
    Model & Layout & $\rho-1$ & $\Delta$Remote & $\Delta\sum$Peak & Eff. \\
    \midrule
    Tiny & Baseline           & 72.7\% & --     & --    & 76.7\% \\
         & Reroute     & 27.0\% & +38.3\% & +8.6\% & 89.0\% \\
         & Full  & 12.7\% & +38.3\% & +1.3\% & 93.0\% \\
    \midrule
    Flash & Baseline           & 40.1\% & --     & --    & 85.1\% \\
          & Reroute     & 13.7\% & +23.3\% & +7.6\% & 93.3\% \\
          & Full &  9.4\% & +23.3\% & +4.9\% & 94.9\% \\
    \bottomrule
  \end{tabular}
\end{table}

\subsection{Device-Side Analysis of Online Throughput}
\label{sec:trace-attribution}

To examine the Tiny 16K H20-3e throughput result in
Figure~\ref{fig:training-throughput}, we collect PyTorch traces from one randomly
selected sealed optimizer window under each of Baseline, Reroute, and Full. We join
kernel external IDs to their launching operators,
form rank-local interval unions, and normalize by each run's observed training
tokens. The aggregate attention, dense/linear, expert, and routing compute
varies by only 0.12\% across traces; Table~\ref{tab:trace-breakdown} groups
these terms as Compute.

Relative to Baseline, Reroute has 5.9\% more DeepEP payload and 10.7\% more
dispatch-side wait, but 74.3\% less combine-side wait. Full has 3.3\% more
payload, 32.1\% less dispatch-side wait, 82.1\% less combine-side wait, and
9.4\% lower total profiled rank-GPU cost. Chronologically aligned combine
calls attribute 89.0--92.4\% of combine wait to expert-stage rank tails.
Full \sysname{} has an expert-tail term \textbf{81.4\%} below Baseline.
The remaining expert tail is only
$2.971\,\mu$s/token, or 1.95\% of the current profiled total, indicating limited
headroom from further tightening expert balance.

Full \sysname{} lowers combined D-wait and
C-wait from $26.496$ to $9.010\,\mu$s/training token, a 66.0\% reduction. The
Tiny 8K replay in Section~\ref{sec:mechanism-decomposition} reduces the modeled
peak-minus-mean joint gap (hatched in Figure~\ref{fig:load-breakdown}) by
79.6\% over 20 windows; the replay isolates compute gaps, whereas the
single-window trace also includes communication and synchronization, which can
dilute the reduction. Appendix~H details the operator and wait
decomposition.

\begin{table}[t]
  \centering
  \caption{Tiny 16K profiled rank-GPU time ($\mu$s/training token).}
  \label{tab:trace-breakdown}
  \RoutePackTableFont
  \setlength{\tabcolsep}{1.4pt}
  \begin{tabular*}{\columnwidth}{@{\extracolsep{\fill}}lrrrrrr@{}}
    \toprule
    Layout & Compute & Payload & D-wait & C-wait & Other & Total \\
    \midrule
    Baseline & 109.7 & \textbf{13.6} & 8.5 & 18.0 & \textbf{18.2} & 168.0 \\
    Reroute  & 109.7 & 14.4 & 9.4 &  4.6 & 19.4 & 157.6 \\
    Full     & 109.8 & 14.1 & \textbf{5.8} & \textbf{3.2} & 19.4 & \textbf{152.3} \\
    \bottomrule
  \end{tabular*}
  \vspace{-2pt}
  \begin{minipage}{\columnwidth}
    \footnotesize D/C-wait estimate differential waiting inside DeepEP
    dispatch/combine notify kernels. Values characterize active GPU capacity,
    exclude idle gaps, and are not per-token latency.
  \end{minipage}
\end{table}

\subsection{Deployment Implications}

\sysname{} assumes fresh routing replay for each optimizer window. Without it,
expert-aware planning must use estimated demand: feasibility is preserved, but
balance gains are not guaranteed. Search budget trades plan quality for host
cost; trajectory bijection and the step-boundary placement contract preserve
training correctness.

\subsection{Limitations and Future Work}
\label{sec:limitations}

Broader validation should test workload and implementation generality
separately: longer GRPO runs on varied task mixtures and group sizes (do the
gains survive changing routing and reward dynamics?), Qwen- and GLM-family MoE
models with other attention and routing designs~\cite{qwen3next,glm45}, and an
open-source RL framework such as Slime~\cite{slimepacking}, exercising replay
capture, placement-consistent state loading, and trajectory rescheduling
outside our in-house admission path. More windows and GPU traces would
characterize D-/C-wait and expert-tail variation, and matched seal-to-update
timing across profile acquisition, placement commit, and state loading would
test full-window gains.

Tiny's multi-second post-state batch await remains an admission straggler;
instrumenting and parallelizing worker preparation and handoffs could remove it.

Calibration of grouped-GEMM latency across microbatch token shapes,
active-expert counts, and tile occupancy would sharpen the rank-load proxy.
Communication-aware planning could then incorporate DeepEP's deduplicated
fanout and endpoint peaks, with weights from topology-specific startup and
overlap measurements. The objective does not model pipeline bubbles; in the
PP=4 Flash deployments, PP enters only through the $2\,\mathrm{PP}$ row
floor, and modeling bubbles is a natural extension.

\section{Conclusion}
\label{sec:conclusion}

We presented \sysname, a replay-guided planner coordinating state-consistent
expert placement with whole-trajectory rescheduling. Incremental repair lowers
the window-level physical-rank bound, while parallel population annealing
balances dense-side work and row-local expert tails at a capacity-efficient row
count without changing logical routing or the deployed topology. Across six
deployments, full \sysname{} records 4.44--11.85\% higher
post-admission Megatron training-region TPS than Base and 1.33--5.28\%
higher than rerouting-only within the same region.
The communication analysis exposes a locality--balance trade-off.

\appendix
\section{Routing and Dense-Side Details}
\label{app:routing-attention}

This section complements Sections~2.1--2.2. Before planning,
the replayed token-level expert IDs of each trajectory are summed into its
exact per-layer expert-demand vector $\mathbf A_i$.

The dense-side proxy assigns every operator to one of two
coefficients. Work proportional to the token count---dense FFNs, projections,
recurrent state updates, and attention over a fixed-width window---enters
$\alpha_s$; token-to-token score and value aggregation enters $\beta_s$. Hence
KDA and Gated DeltaNet have $\beta_s=0$, whereas MLA and full Gated Attention
have $\beta_s>0$. MQA and GQA shrink the projection and KV-state constants but
keep the pairwise term~\cite{gqa}, and top-$k$ or compressed attention patterns
can be absorbed into calibrated coefficients.

In the evaluated Ling-3.0 checkpoints, KDA and MLA run as separate
stages~\cite{ling3flash,ling3models}. Dense FFNs and KDA depend only on the
token count, so their coefficients are merged under one DP maximum; MLA keeps
its linear and quadratic terms together inside each cell and takes its own
stage maximum. The result is a compute proxy rather than a general kernel
model; shape-aware calibration and the grouped-GEMM validation
(Section~\ref{app:grouped-gemm-validation}) are treated separately.

\section{EP Communication Model Derivation}
\label{app:communication-model}

Section~6.1 summarizes this diagnostic; here we derive it. An EP
layer sends routed activations to the ranks that own the selected experts
(dispatch) and returns the outputs through a second all-to-allv (combine). As
a reference point, take one execution row $r\in[R]$ of EDP shard $g$ at MoE
layer $l\in\mathcal L_{\mathrm{MoE}}$, with $n=|\mathcal P_g|$ balanced EP
ranks, $b$ trajectories of $t$ tokens each, top-$k$ routing, and activation
width $h_l$. With uniform sources and destinations, each rank moves
\begin{RoutePackCompactDisplay}
\begin{equation}
  Q_{\mathrm{uni}}
  =2\frac{kbt}{n}h_l\frac{n-1}{n}
  =\frac{2kbth_l(n-1)}{n^2}
  \label{eq:uniform-ep-volume}
\end{equation}
\end{RoutePackCompactDisplay}
activation elements across dispatch and combine~\cite{jin2026megascale}. At
$w$ bytes per element and effective endpoint bandwidth $B_l$, one
dispatch--combine pass then takes $wQ_{\mathrm{uni}}/B_l$ when only bandwidth
matters; backward passes and recomputation repeat this traffic and add their
own multiplicative and overlap terms.

Real routing is not uniform. Let $\sigma_g:\mathcal D_g\to\mathcal P_g$ give
the EP rank that hosts each source DP slot of shard $g$. For $u\ne v$, the row
assignment $x_{i,r,d}$, the placement $\pi_l$, and the routed demand
$q_{i,l,v}(\pi_l)$ determine the off-rank traffic matrix
\begin{RoutePackCompactDisplay}
\begin{equation}
  M_{r,g,l}[u,v]=
  \sum_{\substack{d\in\mathcal D_g:\\ \sigma_g(d)=u}}
  \sum_i x_{i,r,d}q_{i,l,v}(\pi_l).
  \label{eq:dispatch-matrix}
\end{equation}
\end{RoutePackCompactDisplay}
Local traffic is excluded, so the diagonal is zero, and combine traverses
$M_{r,g,l}^{\mathsf T}$. Dropping the index $g$, rank $u$ sends
$O_{r,l,u}=\sum_v M_{r,l}[u,v]$ and rank $v$ receives
$I_{r,l,v}=\sum_u M_{r,l}[u,v]$. The total off-rank volume is
$V_{r,l}=\sum_{u,v}M_{r,l}[u,v]$, and the busiest endpoint carries
\begin{RoutePackCompactDisplay}
\begin{equation}
  m_{r,l}=\max\!\left\{\max_u O_{r,l,u},\;\max_v I_{r,l,v}\right\}.
  \label{eq:dispatch-endpoint-peak}
\end{equation}
\end{RoutePackCompactDisplay}

SABRE bounds the completion time of an exchange by its busiest
endpoint~\cite{wei2026sabre}:
\begin{RoutePackCompactDisplay}
\begin{equation}
  T\geq\max_p\left\{\frac{\mathrm{send}_p}{B_p^s},
                       \frac{\mathrm{recv}_p}{B_p^r}\right\}.
  \label{eq:sabre}
\end{equation}
\end{RoutePackCompactDisplay}
With homogeneous full-duplex endpoints and both dispatch and combine counted,
this gives a per-row, per-layer service estimate and its balanced counterpart:
\begin{RoutePackDenseDisplay}
\begin{align}
  \widehat T^{\mathrm{comm}}_{r,l}
    &=\frac{2wh_l}{B_l}m_{r,l},
  &\widehat T^{\mathrm{LB}}_{r,l}
    &=\frac{2wh_l}{B_l}\frac{V_{r,l}}{n}, \\
  \rho_{r,l}
    &=\frac{nm_{r,l}}{V_{r,l}},
  &\Delta\widehat T^{\mathrm{imb}}_{r,l}
    &=\widehat T^{\mathrm{LB}}_{r,l}(\rho_{r,l}-1).
  \label{eq:dispatch-time-model}
\end{align}
\end{RoutePackDenseDisplay}
Since $m_{r,l}\geq V_{r,l}/n$, the balanced term is the fastest endpoint
service possible for the given off-rank volume, and $\rho_{r,l}-1$ measures the
excess. Uniform routing gives $V_{r,l}=kbt(n-1)/n$ and $m_{r,l}=V_{r,l}/n$, so
$\widehat T^{\mathrm{comm}}_{r,l}$ reduces to $wQ_{\mathrm{uni}}/B_l$ as
expected. The send/receive maximum remains a capacity lower bound even when
overlap is imperfect; reaching it would additionally require concurrent
service at every endpoint and no narrower bottleneck elsewhere in the topology.

\section{Grouped-GEMM Cost-Model Validation}
\label{app:grouped-gemm-validation}

This section gives the protocol behind Section~2.3.1 and Figure~4.
The shapes are per-expert token counts recorded at the TEGroupedMLP input after
DeepEP dispatch: 480 layer--EP-group observations with 16 ranks per group and
32 local experts per rank, i.e., 7,680 rank shapes. For each shape we time bf16
forward and backward passes of the $2560\rightarrow1536$ and
$768\rightarrow2560$ projections, with five warm-up and 20 timed iterations.
Rank latency has a median of 3.8004~ms, a 90th percentile of 13.5175~ms, a 99th
percentile of 24.4649~ms, and a maximum of 38.0371~ms. Rank demand has a median
of 34,014 tokens, a 90th percentile of 163,038, and a maximum of 459,985. The
benchmark covers grouped-GEMM compute only; communication is reflected in the
trainer-level throughput measurements.

Over the 480 layer--EP-group observations, the EP-rank max/mean ratio of token
counts and of latency is distributed as follows:\par
\noindent{\RoutePackTableFont
\setlength{\tabcolsep}{2.2pt}
\begin{tabular}{@{}lrrrrrr@{}}
  \toprule
  & Mean & P50 & P90 & P95 & P99 & Max \\
  \midrule
  Token & 3.5698 & 3.5244 & 5.0788 & 5.3517 & 6.7437 & 7.0188 \\
  Latency & 2.9049 & 2.8099 & 3.9500 & 4.1629 & 5.2246 & 5.3572 \\
  \bottomrule
\end{tabular}}\par
The single-rank fixed-token control in Table~2 complements this
sweep.

\section{State-Consistent Expert Placement}
\label{app:state-consistency}

\begin{routealgorithm}{Incremental layer-local expert placement}
  \label{alg:lpt-placement}
  \AlgRequire{Aggregate loads $\{L_{l,e}\}_{e=1}^{E}$; prior map $\pi_l^{-}$;
  physical ranks $\mathcal P$; $H=E/|\mathcal P|$ slots per rank; CV gate
  $\epsilon$}
  \AlgEnsure{Slot permutation $\pi_l$ with $H$ experts per rank}
  \AlgLine{$\pi_l\gets\pi_l^{-}$; compute $G_{l,p}(\pi_l)$ for all $p$}
  \AlgLine{While $\operatorname{CV}(G_{l,:})>\epsilon$, find the cross-rank
    pair $(e_a,e_b)$ whose swap gives the largest strict reduction in
    $\sum_pG_{l,p}^2$}
  \AlgLine{If no improving pair exists, stop; otherwise swap their rank
    assignments and update the two affected rank loads}
  \AlgLine{Keep the exact slots of experts that remain on their source rank;
    place incoming experts into slots vacated by outgoing experts}
  \AlgLine{\textbf{return} $\pi_l$}
\end{routealgorithm}

Algorithm~\ref{alg:lpt-placement} is the incremental repair of
Section~3.2. Each resulting $\pi_l$ takes effect only at a training-step
boundary. Expert FC weights are copied into the existing destination tensors,
so parameter objects and gradient-buffer descriptors are never rebuilt. The
same map moves every expert-indexed companion: master weights, optimizer
moments, quantization scales, and other per-expert metadata. Under a
distributed optimizer, each source gbuf range is sent within its
state-consistency domain and written into the existing destination range,
keeping ownership and tensor addresses fixed.

The router still indexes experts logically; only the dispatcher translates a
replayed top-$k$ expert ID $e$ into physical slot $\pi_l(e)$. Routing
probabilities therefore keep their original differentiable columns, and expert
gradients accumulate in destination-slot gbuf views whose optimizer state
moved with them. DeepEP receives the remapped physical indices through its
unchanged dispatch/combine API~\cite{deepep}, and a checkpoint can be exported
in canonical order by applying $\pi_l^{-1}$ at a step boundary.

Every TP shard applies the same permutation, and DP replicas and EDP lanes
commit it within their state-consistency domain. Because the layout stays
fixed for the whole forward--backward interval, TP, SP, CP, PP, 1F1B,
activation recomputation, gradient accumulation, and distributed-optimizer
sharding all see unchanged shapes, groups, and dependencies. Mechanisms that
add replicas or change ownership within an interval fall outside this
contract.

\section{Trajectory Rescheduling Search Details}
\label{app:packing-search}

\subsection{Diverse Fixed-Row Seeding}

This subsection details the seeding outlined in Section~4.1. For
target $R^\star$, the planner creates $K=R^\star D$ cells and always keeps two
seeds: the canonical FFD layout and its EDP-aware pairing. Further trials
perturb the nonincreasing length order (\opWindowShuffle), let
\opRandomizedBestFit choose among the most occupied cells that still fit, and
then pair the completed cells by their incremental EDP-rank peaks. Any trial
that leaves $\mathcal X_{R^\star}$ is discarded.

Layouts that differ only in trajectory order within a cell, in row order, or
in the order of exchangeable DP slots inside an EDP shard are identical, and
canonicalization removes these symmetries. To measure structural diversity,
the planner caches $\phi_k(x)=\operatorname{sort}(H(x)u_k)$ for the normalized
row-feature matrix $H(x)$ and fixed projections $u_k$, and combines the
deterministic sliced-Wasserstein term
$q^{-1}\sum_k\|\phi_k(x)-\phi_k(y)\|_1/R^\star$ with cell- and row-level
co-assignment~\cite{rabin2011slicedwasserstein}. Selection starts from the
exact-score winner and then trades distance from the already selected set
against the normalized gap in primary score.

\begin{routealgorithm}{Diverse fixed-row initialization}
  \label{alg:seed-population}
  \AlgRequire{Trajectories; $C$; $R^\star$; $D$; EDP groups; fixed $\pi$;
  candidate and population budgets}
  \AlgEnsure{A feasible initial population $\mathcal P_0$}
  \AlgLine{$x_{\mathrm{length}}\gets\textsc{FixedRowFFD}(\mathcal S,C,R^\star D)$}
  \AlgLine{$x_{\mathrm{paired}}\gets\opEDPAwarePair(x_{\mathrm{length}},\pi)$;
    insert both into canonical map $\mathcal C$}
  \AlgLine{Generate randomized best-fit fixed-row trials and reject any trial
    outside $\mathcal X_{R^\star}$}
  \AlgLine{Pair trial cells with \opEDPAwarePair and retain the best exact
    score per canonical signature}
  \AlgLine{$\mathcal P_0\gets\opQualityDiversitySelect(\mathcal C)$; repeat
    selected members only if needed to fill the worker population}
  \AlgLine{\textbf{return} $\mathcal P_0$}
\end{routealgorithm}

\subsection{Cooling, Path Weights, and Resampling}

Population-SA splits a geometric schedule $T_0>\cdots>T_N$ into segments;
inside segment $n$, step $k$ runs at

\begin{RoutePackCompactDisplay}
\begin{equation}
  T_{n,k}=T_n\left(\frac{T_{n+1}}{T_n}\right)^{k/K}.
  \label{eq:pa-chain-temperature}
\end{equation}
\end{RoutePackCompactDisplay}

Metropolis acceptance compares the scalar energy

\begin{RoutePackCompactDisplay}
\begin{equation}
  \mathcal E(x)=s_0(x)+10^{-6}s_1(x)+10^{-9}s_2(x),
  \label{eq:pa-energy}
\end{equation}
\end{RoutePackCompactDisplay}

where $(s_0,s_1,s_2)=\operatorname{Score}(x;\pi)$. The scalarization serves
only acceptance; winners are always chosen by the exact tuple. Measured
against a reference energy $E_{\mathrm{ref}}$ and scale $E_{\mathrm{scale}}$,
worker $j$ accumulates the path weight

\begin{RoutePackCompactDisplay}
\begin{equation}
  \ell_j=-\sum_{k=1}^{K}(T_{n,k}^{-1}-T_{n,k-1}^{-1})
  \frac{\mathcal E(y_{j,k-1})-E_{\mathrm{ref}}}{E_{\mathrm{scale}}}.
  \label{eq:pa-path-weight}
\end{equation}
\end{RoutePackCompactDisplay}

At each segment boundary, the planner turns the $\ell_j$ into resampling
probabilities and flattens them with the smallest tempering factor whose
multinomial proxy yields the requested number of distinct parents. The
strongest distinct parents are each kept once, and systematic resampling fills
the remaining positions. Terminal states seed the next segment, while each
chain's best state feeds a separate exact-score incumbent, so a good transient
solution is never lost even though it does not replace the parent state tied
to the path weight.

\begin{routealgorithmfloat}{Fixed-row Population Annealing}
  \label{alg:population-sa}
  \AlgRequire{Initial population $\mathcal P_0$; geometric temperatures;
  segment length $K$; fixed $(R^\star,\pi)$}
  \AlgEnsure{A valid, non-regressing layout $x^\star$}
  \AlgLine{Initialize $x^\star$ from the best exact-score seed}
  \AlgLine{For each temperature segment, run one guided swap/relocation chain
    per population member in parallel}
  \AlgLine{Accept feasible proposals by Metropolis energy and update each
    per-chain best state}
  \AlgLine{Update $x^\star$ by exact lexicographic comparison}
  \AlgLine{Temper path weights, protect strong distinct parents, and
    systematically resample terminal states}
  \AlgLine{Validate coverage, capacity, nonempty cells, row count, topology,
    and non-regression; \textbf{return} $x^\star$}
\end{routealgorithmfloat}

\section{Planner Complexity}
\label{app:complexity}

\begin{figure*}[!t]
  \centering
  \makebox[\textwidth][c]{%
  \includegraphics[width=1.02\textwidth]{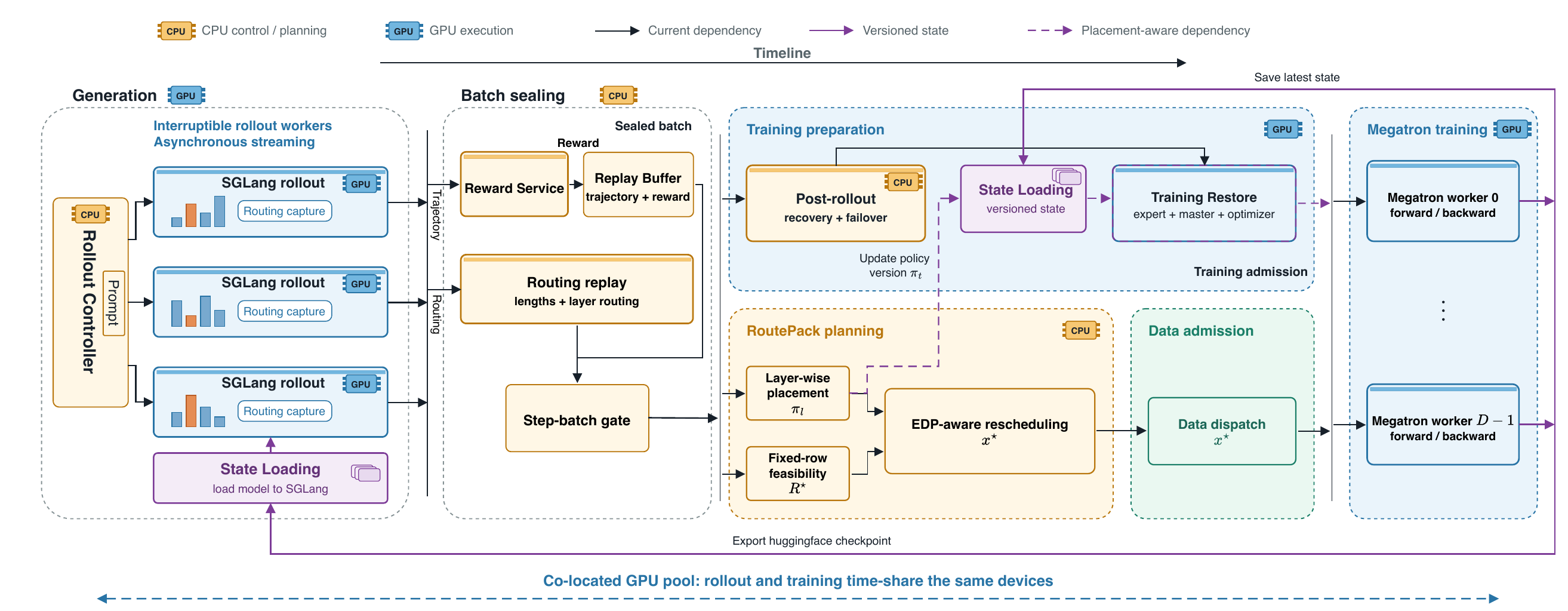}
  }
  \Description{SGLang rollout workers stream trajectories to reward evaluation and
  routing replay. A CPU planner computes row count, layer-wise expert placement,
  and EDP-local data layout. Model-state preparation and CPU rescheduling execute in
  parallel and join before Megatron training admission.}
  \caption{\sysname{} in the RL framework: CPU planning overlaps post-rollout
  processing and state preparation.}
  \label{fig:framework-integration-supplement}
\end{figure*}

Figure~\ref{fig:framework-integration-supplement} shows where the planner sits
in the framework timeline; Figure~9 gives the corresponding target
schedule.

Let $n=|\mathcal S|$ trajectories fill $h=R^\star D$ cells, and let
$L,E,P,G$ be the numbers of MoE layers, logical experts, physical EP ranks, and
EDP shards. Aggregating routes and repairing placement costs
$O(A_{\mathrm{route}}+L(E\log E+EP))$. With $W$ workers and fixed population,
temperature, and proposal budgets, annealing takes at most

\begin{RoutePackCompactDisplay}
\begin{equation}
  O\!\left(N_T\left\lceil\frac{P_{\mathrm{pop}}}{W}\right\rceil
  K(R^\star GLP+R^\star D)\right)
  \label{eq:planner-complexity}
\end{equation}
\end{RoutePackCompactDisplay}

wall time. Randomized best-fit costs $O(nh\log h)$ and cell pairing
$O(h^2GLP)$; with cached $q$-projection sketches, encoding a candidate costs
$O(MqR^\star(F+\log R^\star))$ and comparing two profiles $O(qR^\star)$. When
enough workers saturate candidate generation and population transitions, the
total is

\begin{RoutePackDenseDisplay}
\begin{equation}
\begin{aligned}
T_{\mathrm{CPU}}^{\mathrm{sat}}=O\bigl(&A_{\mathrm{route}}+L(E\log E+EP)
 +nh\log h+h^2GLP\\
 &+qR^\star(F+\log R^\star)
 +P_{\mathrm{pop}}(n+qR^\star)\\
 &+N_TK(R^\star GLP+R^\star D)\bigr).
\end{aligned}
\label{eq:saturated-planner-complexity}
\end{equation}
\end{RoutePackDenseDisplay}

For a fixed topology and search budget, every added search term is polynomial
in the row count. These bounds are conservative; Section~5.3
reports the measured planning time and exposed wait.

\section{Online Measurement Details}
\label{app:online-details}

\begin{figure*}[!t]
  \centering
  \includegraphics[width=\textwidth]{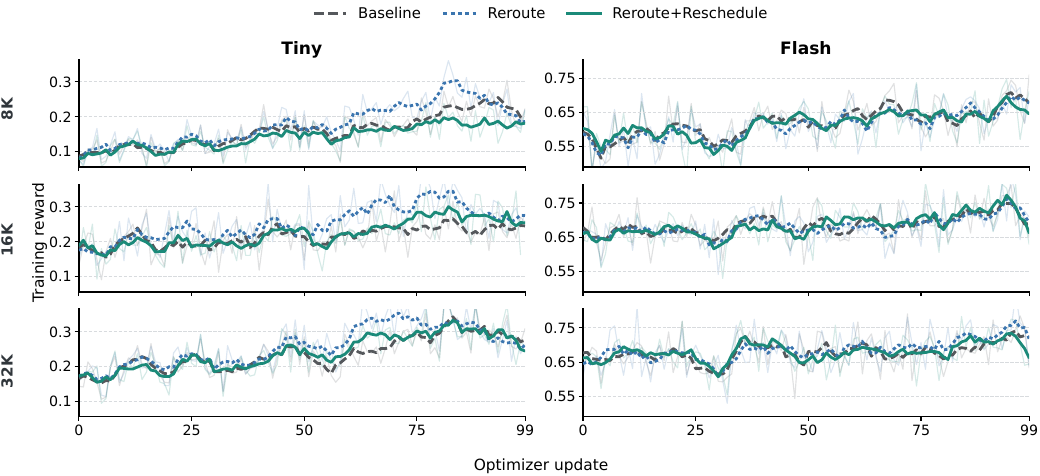}
  \Description{Six reward plots arranged by context window and model compare
  Baseline, Expert reroute, and Reroute plus trajectory rescheduling over the
  first 100 optimizer updates. All curves maintain an increasing training
  trend.}
  \caption{Training reward over the first 100 optimizer updates on six
  H20-3e configurations. Faint lines show per-update rewards; emphasized traces
  show five-update moving averages.}
  \label{fig:reward-curves}
\end{figure*}

Table~\ref{tab:app-training-throughput} lists the absolute rates behind
Figure~10, and Figure~\ref{fig:reward-curves} shows the reward
curves of the same runs.

\paragraph{Choice of online policies.}
The closest expert-side systems, ReLibra and ForeMoE, are not open
source~\cite{relibra,foremoe}. Their microbatch-level expert replication and
token splitting also change the training unit that our state-consistency
contract keeps fixed, so a like-for-like deployment is not possible;
Table~4 compares their signals and controls instead.
Table~6 isolates the reverse control order (rescheduling under identity
placement), and Section~\ref{app:offline-packing} shows that global length-only
FFD raises the joint proxy on all 20 Tiny windows and on 17 of 20 Flash
windows. The online study therefore follows the deployed hierarchy: Base,
Reroute, and Full.

\begin{table}[t]
  \centering
  \caption{Megatron training-region throughput (kTok/s); parentheses give
  within-run per-update TPS CV (\%).}
  \label{tab:app-training-throughput}
  \RoutePackTableFont
  \setlength{\tabcolsep}{3pt}
  \begin{tabular*}{\columnwidth}{@{\extracolsep{\fill}}lrrr@{}}
    \toprule
    Run & Base & Reroute & +Resched. \\
    \midrule
    Flash 8K & 37.005 (3.65) & 37.994 (4.14) & 38.649 (4.04) \\
    Flash 16K & 57.775 (4.92) & 59.416 (5.22) & 61.175 (5.95) \\
    Flash 32K & 44.041 (5.46) & 46.791 (5.98) & 49.260 (5.34) \\
    Tiny 8K & 49.504 (2.40) & 51.051 (1.56) & 52.225 (1.31) \\
    Tiny 16K & 47.119 (2.38) & 49.286 (1.64) & 51.064 (1.33) \\
    Tiny 32K & 78.323 (2.12) & 81.312 (2.42) & 82.391 (2.23) \\
    \bottomrule
  \end{tabular*}
\end{table}

Tiny 16K uses 98 post-warm-up updates per policy, and the other H20-3e runs
use 99. Because each policy is a separate stochastic job, update $i$ of one
policy does not see the same trajectories as update $i$ of another. The
fixed-input replay of Section~4.2 shows the same Base--Reroute--Full
ordering in modeled joint cost on shared windows. The CVs in
Table~\ref{tab:app-training-throughput} describe variation within one online
job, not uncertainty across independent runs.

\section{Profiler Accounting}
\label{app:profiler-accounting}

The case study of Section~6.2 covers all eight ranks of one randomly
chosen sealed window per method. We keep Kineto's GPU kernel, memcpy, and
memset events, with timestamps and durations in microseconds. Writing
$|\bigcup\mathcal I|$ for the interval-union duration of an event set
$\mathcal I$, overlapping events of one category are counted once and nested
host annotations are never summed. All methods are classified by the same
ordered rules (Table~\ref{tab:app-trace-taxonomy}); a kernel with an ambiguous
name inherits the category of the CPU operator that launched it, found through
Kineto's external ID.

\begin{table}[t]
  \centering
  \caption{GPU trace category rules.}
  \label{tab:app-trace-taxonomy}
  \RoutePackTableFont
  \setlength{\tabcolsep}{3pt}
  \begin{tabular}{@{}>{\raggedright\arraybackslash}p{0.20\columnwidth}
                      >{\raggedright\arraybackslash}p{0.74\columnwidth}@{}}
    \toprule
    Category & Included device events \\
    \midrule
    Attention & KDA and cuDNN SDPA/Flash-attention kernels, plus associated
      attention BMM, mask, and reduction operators recovered by external ID. \\
    Dense/linear & Generic GEMM kernels launched by \texttt{mm},
      \texttt{addmm}, \texttt{linear}, or \texttt{matmul}. \\
    Expert & Grouped-GEMM kernels (\texttt{grouped}, \texttt{group\_gemm}, or
      \texttt{nvjet}) and kernels launched by grouped-linear backward. \\
    Routing & Top-$k$, token index/scatter-add, and router mask/reduction
      operators recovered by external ID. \\
    EP payload & DeepEP dispatch and combine kernels whose names do not contain
      \texttt{notify}. \\
    Dispatch wait & Cross-rank duration above the per-call minimum for
      uncached- and cached-dispatch notify kernels. \\
    Combine wait & Cross-rank duration above the per-call minimum for
      cached-combine notify kernels. \\
    Other & Residual GPU-union time: model glue, non-EP collectives, copies and
      sets, optimizer and unmatched kernels, and the common notify/control floor. \\
    \bottomrule
  \end{tabular}
\end{table}

D-wait and C-wait in Table~12 are the Dispatch-wait and
Combine-wait rows here. The names give the synchronization point, not the
cause: skew in attention, dense, or routing work can surface as dispatch-side
waiting, and payload or expert-path skew as combine-side waiting.

Both waits come from DeepEP's notify kernels for uncached dispatch, cached
dispatch, and cached combine. Within each kind, calls are ordered by GPU
timestamp and matched across ranks by ordinal. With $d_{r,k,j}$ the duration
of call $j$ of kind $k$ on rank $r$, the wait is
\begin{RoutePackCompactDisplay}
\begin{equation}
  T_{\mathrm{barrier}}=
  \sum_{k,j}\sum_{r=0}^{7}
  \left(d_{r,k,j}-\min_{q\in[8]}d_{q,k,j}\right).
  \label{eq:trace-barrier-accounting}
\end{equation}
\end{RoutePackCompactDisplay}
Subtracting the per-call minimum removes the control work that every rank
performs, so the remainder estimates how long ranks wait for one another, not
network latency or total notify time. Every rank issues the same number of
calls of each kind within a run, so ordinal matching drops nothing: Baseline
and Reroute issue 1,334 dispatch, 1,334 cached-dispatch, and 2,668
cached-combine calls per rank, and Full issues 1,288, 1,288, and 2,576.

Let $T_{m,r}^{\mathrm{all}}$ be the union of all retained GPU intervals of
method $m$ on rank $r$. \textbf{Other} is what remains of
$\sum_rT_{m,r}^{\mathrm{all}}$ after subtracting the compute categories, EP
payload, and the two waits; attention auxiliaries and recognizable dense,
linear, and grouped-linear kernels are thus charged to their model component
rather than left in Other. On every rank, the category unions overlap by at
most 0.085\% of the all-GPU union, and this overlap stays in Other together
with unmatched events. An attention- or dense-side straggler appears only if it
materializes as a retained GPU collective or control kernel; idle GPU gaps and
host-side synchronization lie outside the interval union.

With $N_m=512\bar\ell_m$ observed training tokens, each entry is
\begin{RoutePackCompactDisplay}
\begin{equation}
  C_{m,c}=\frac{\sum_{r=0}^{7}T_{m,r,c}}{N_m}
  \quad [\mu\mathrm{s}/\text{training token}],
  \label{eq:trace-token-normalization}
\end{equation}
\end{RoutePackCompactDisplay}
so by construction Total equals
$\sum_c C_{m,c}=\sum_rT_{m,r}^{\mathrm{all}}/N_m$. The unit is consumed
rank-GPU capacity per training token; it is neither per-token latency nor
unprofiled end-to-end time. Normalizing by tokens softens the effect of
independent rollout sampling, but the result remains a mechanism case study
rather than a paired throughput comparison.

\begin{table}[t]
  \centering
  \caption{Disaggregated Tiny 16K trace accounting ($\mu$s/training token).}
  \label{tab:app-trace-breakdown}
  \RoutePackTableFont
  \setlength{\tabcolsep}{3pt}
  \begin{tabular*}{\columnwidth}{@{\extracolsep{\fill}}lrrr@{}}
    \toprule
    Category & Baseline & Reroute & Full \\
    \midrule
    Attention       & 44.875 & 44.793 & 44.982 \\
    Dense/linear    &  0.433 &  0.433 &  0.432 \\
    Expert          & 57.270 & 57.388 & 57.292 \\
    Routing         &  7.113 &  7.126 &  7.116 \\
    EP payload      & 13.619 & 14.416 & 14.073 \\
    Dispatch wait   &  8.532 &  9.442 &  5.795 \\
    Combine wait    & 17.964 &  4.616 &  3.215 \\
    Other           & 18.183 & 19.384 & 19.359 \\
    \midrule
    Total           &167.989 &157.598 &152.264 \\
    \bottomrule
  \end{tabular*}
\end{table}

The first four rows of Table~\ref{tab:app-trace-breakdown} add up to the
Compute column of Table~12; the remaining rows map one-to-one to
its other columns.

\paragraph{Dispatch-side decomposition.}
Dispatch wait spans two synchronization points: the uncached notify exchanges
each row's counts and metadata, and the cached notify reuses that layout.
Rerouting raises both slightly, whereas joint rescheduling lowers both
(Table~\ref{tab:app-dispatch-wait}).

\begin{table}[t]
  \centering
  \caption{Dispatch-wait decomposition ($\mu$s/training token).}
  \label{tab:app-dispatch-wait}
  \RoutePackTableFont
  \setlength{\tabcolsep}{3pt}
  \begin{tabular*}{\columnwidth}{@{\extracolsep{\fill}}lrrr@{}}
    \toprule
    Method & Uncached & Cached & Total \\
    \midrule
    Baseline & 4.397 & 4.134 & 8.532 \\
    Reroute  & 4.711 & 4.731 & 9.442 \\
    Full     & 3.372 & 2.423 & 5.795 \\
    \bottomrule
  \end{tabular*}
\end{table}

\paragraph{Combine-side attribution.}
For combine call $j$, $g_{r,j}$ spans rank $r$'s first grouped GEMM after the
preceding dispatch payload through its last grouped GEMM before the combine
notify, and the expert tail is $\sum_j\sum_r(\max_q g_{q,j}-g_{r,j})$. Each
combine matches exactly one dispatch, and the median spread of notify
completion across the eight ranks is only $41\,\mu$s. The expert tail makes up
89.0--92.4\% of combine wait (Table~\ref{tab:app-combine-attribution});
Reroute shrinks it by 73.6\% and Full by 81.4\%, which accounts for 88.1\% and
88.2\% of their respective combine-wait reductions.

\begin{table}[H]
  \centering
  \caption{Combine-wait attribution ($\mu$s/training token).}
  \label{tab:app-combine-attribution}
  \RoutePackTableFont
  \setlength{\tabcolsep}{3pt}
  \begin{tabular*}{\columnwidth}{@{\extracolsep{\fill}}lrrrr@{}}
    \toprule
    Method & Expert tail & Residual & Total & Share \\
    \midrule
    Baseline & 15.987 & 1.977 & 17.964 & 89.0\% \\
    Reroute  &  4.229 & 0.387 &  4.616 & 91.6\% \\
    Full     &  2.971 & 0.244 &  3.215 & 92.4\% \\
    \bottomrule
  \end{tabular*}
\end{table}

\paragraph{Other.}
Table~\ref{tab:app-other-breakdown} splits the residual. Model glue collects
the remaining operator-launched elementwise, copy, normalization, residual,
and miscellaneous kernels that cannot be safely charged to one model
component, and the negative row corrects the small overlap between category
unions. Nearly all of Other is stable framework and model glue, not hidden
dense or attention GEMM work.

\begin{table}[H]
  \centering
  \caption{Residual Other decomposition ($\mu$s/training token).}
  \label{tab:app-other-breakdown}
  \RoutePackTableFont
  \setlength{\tabcolsep}{2.5pt}
  \begin{tabular*}{\columnwidth}{@{\extracolsep{\fill}}lrrr@{}}
    \toprule
    Component & Baseline & Reroute & Full \\
    \midrule
    Model glue           & 17.614 & 18.782 & 18.764 \\
    Non-EP comm./memory  &  0.305 &  0.340 &  0.336 \\
    Notify/control floor &  0.276 &  0.273 &  0.274 \\
    Unmatched GEMM       &  0.110 &  0.111 &  0.108 \\
    Overlap adjustment   & $-0.122$ & $-0.122$ & $-0.123$ \\
    \midrule
    Total                & 18.183 & 19.384 & 19.359 \\
    \bottomrule
  \end{tabular*}
\end{table}

\section{Online Plan Refinement Details}
\label{app:online-plan-refinement}

Table~\ref{tab:app-online-plan-refinement} lists the exact per-configuration
means plotted in Figure~11.

\begin{table}[H]
  \centering
  \caption{Online plan refinement relative to the fractional lower bound.}
  \label{tab:app-online-plan-refinement}
  \RoutePackTableFont
  \setlength{\tabcolsep}{3pt}
  \begin{tabular*}{\columnwidth}{@{\extracolsep{\fill}}lrrrr@{}}
    \toprule
    Run & $J_0/B$ & $J_1/B$ & $\eta$ (\%) & $\Delta J$ (\%) \\
    \midrule
    Flash 8K  & 1.0758 & 1.0529 & 28.91 & 2.11 \\
    Flash 16K & 1.0852 & 1.0553 & 33.58 & 2.75 \\
    Flash 32K & 1.0612 & 1.0265 & 52.22 & 3.23 \\
    Tiny 8K   & 1.1213 & 1.0643 & 46.03 & 5.06 \\
    Tiny 16K  & 1.0871 & 1.0557 & 32.92 & 2.86 \\
    Tiny 32K  & 1.0774 & 1.0338 & 54.28 & 4.02 \\
    \bottomrule
  \end{tabular*}
\end{table}

\section{Fixed-Row ILP Scaling}
\label{app:ilp-scaling}

Section~4.3 uses five warm profiles per model, drawn with NumPy seed
42: Tiny steps 2, 8, 9, 12, and 13, and Flash steps 2, 9, 14, 15, and 17. Every
nested prefix consists of complete eight-trajectory GRPO groups. For each
subproblem, we first replay the chronological incremental-placement chain and
then freeze the resulting physical EP loads, the global length-only (FFD) row
count, capacity 8,192, DP=8, and PP=1; both models use a single eight-slot EDP
shard.

Population-SA and the ILP backend, Google OR-Tools 9.15 CP-SAT (a
constraint-programming solver built on satisfiability methods), optimize the
same primary objective with at most 64 workers each. Population-SA uses 64
members, two temperature levels, and 10,000 proposals per member and level.
CP-SAT starts from the global FFD plan, supplied as a complete hint and a safe
objective cut, runs for at most 60~s including presolve, and scales weights to
fixed point by $10^3$. The hint does not register as an incumbent, and only an
\textsc{Optimal} status certifies the quantized fixed-row subproblem. Expert
placement, row count, capacity, and EDP topology stay fixed for both methods.

\begin{table}[H]
  \centering
  \caption{Median objective reduction from FFD over five profiles (\%). ILP
  uses OR-Tools CP-SAT. O/F/U: \textsc{Optimal}/\textsc{Feasible}/\textsc{Unknown};
  U has no returned incumbent.}
  \label{tab:app-ilp-scaling}
  \RoutePackTableFont
  \setlength{\tabcolsep}{3pt}
  \begin{tabular*}{\columnwidth}{@{\extracolsep{\fill}}lrrrr@{}}
    \toprule
    Model & $N$ & Pop.-SA & ILP & Status \\
    \midrule
    Tiny  &  32 &  6.63 &  6.64 & 1O+4F \\
    Tiny  &  64 &  5.68 &  5.35 & 5F \\
    Tiny  & 128 &  7.53 &  1.04 & 5F \\
    Tiny  & 256 &  6.19 &    -- & 5U \\
    Flash &  32 & 12.02 & 11.59 & 5F \\
    Flash &  64 &  8.38 &  6.77 & 5F \\
    Flash & 128 &  6.73 &  2.38 & 5F \\
    Flash & 256 &  4.99 &    -- & 5U \\
    \bottomrule
  \end{tabular*}
\end{table}

We report the feasible objective reduction instead of a fraction of the
optimum: within the time limit, the CP-SAT lower bound can remain far from its
incumbent, and \textsc{Unknown} is not evidence of infeasibility. Nor does the
comparison establish global optimality for Population-SA. It shows only that
the bounded production search keeps returning improving feasible layouts as
the general-purpose formulation becomes hard to even initialize within the
available budget.

\section{Offline Trajectory Rescheduling Ablation}
\label{app:offline-packing}

\paragraph{Protocol.}
This ablation uses the 20-window corpus of Section~2.3.2 (Table~3)
with DP=8, EP=8, EDP=1, all other parallel dimensions set to one, and an
8,192-token cell capacity; recorded trajectory lengths decide capacity
feasibility. Flash's recorded PP=4 only imposes an eight-row minimum, which
never binds because every replayed layout has at least 11 rows, so no pipeline
schedule is modeled. Within a window, all three schedules share one
thresholded layer-wise placement that starts from identity and stops at a CV of
$5\times10^{-4}$. The study isolates data layout and does not evaluate
continuous placement or state migration.

\textbf{Baseline} is the in-house DP-local FFD defined in
Section~4.2. The dumps lack live shard identifiers and RNG state, so we
reconstruct it algorithmically: shard size eight, base shuffle seed 42, and the
zero-based window ordinal (the plotted window index minus one) as the shuffle
counter. \textbf{Global FFD} packs the whole window by length into a feasible
row layout, and \textbf{\sysname{}} runs diverse seeding, EDP-aware cell
pairing, and Population-SA at that same global row count. Online measurements
follow the deployed Base--Reroute--Full ladder; this fixed-input replay adds
Global FFD to separate the row-count benefit of length-only compaction from its
effect on expert peaks and from the routing-aware refinement that follows.
Both models use the same planner code and search budget: eight candidate
layouts, 64 population members, two levels, 10,000 proposals per member and
level, at least eight surviving parents, and eight local worker processes,
with rescheduling seed 0 and SA seed 42.

The dense side accumulates two maxima separately, KDA plus dense FFN (linear)
and MLA, with model-specific coefficients (KDA-plus-dense linear, MLA linear,
MLA quadratic) of (148.689, 42.896, 53.333) for Tiny and (155.017, 27.803,
24.889) for Flash. These values are rounded here; the replay uses the supplied
full-precision values without rescaling. All costs are compute proxies, not
measured latency or throughput. Each window's cost is normalized to its own
in-house-baseline reference before averaging, and raw costs are never compared
across models.

\begin{table}[t]
  \centering
  \RoutePackTableFont
  \setlength{\tabcolsep}{3pt}
  \caption{Offline trajectory rescheduling (20 windows per model; Baseline:
  DP-local FFD; costs: mean per-window ratios, baseline=100; Util.: token-cap utilization).}
  \label{tab:offline-packing}
  \begin{tabular}{@{}lrrrrr@{}}
    \toprule
    Schedule & Rows & Util. (\%) & EP & Dense-side & Joint \\
    \midrule
    \multicolumn{6}{c}{MoE layers $\times$ experts: $23\times128$} \\
    Baseline & 63.25 & 85.89 & 100.00 & 100.00 & 100.00 \\
    Global FFD & 60.30 & 90.07 & 113.04 & 93.99 & 103.35 \\
    \sysname{} & 60.30 & 90.07 & 88.41 & 98.66 & 93.61 \\
    \midrule
    \multicolumn{6}{c}{MoE layers $\times$ experts: $40\times512$} \\
    Baseline & 31.70 & 90.36 & 100.00 & 100.00 & 100.00 \\
    Global FFD & 29.95 & 95.77 & 105.84 & 93.19 & 101.04 \\
    \sysname{} & 29.95 & 95.77 & 97.18 & 94.89 & 96.31 \\
    \bottomrule
  \end{tabular}
\end{table}

\begin{figure*}[!t]
  \centering
  \begin{minipage}[t]{0.49\textwidth}
    \centering
    {\small Tiny: $23$ MoE layers, $128$ experts per layer\par}
    \includegraphics[width=\linewidth]{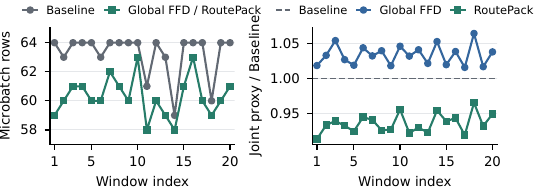}
  \end{minipage}\hfill
  \begin{minipage}[t]{0.49\textwidth}
    \centering
    {\small Flash: $40$ MoE layers, $512$ experts per layer\par}
    \includegraphics[width=\linewidth]{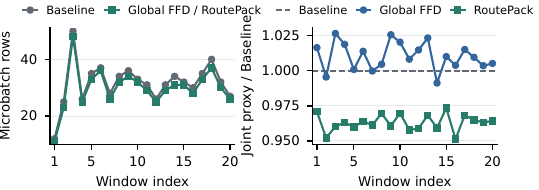}
  \end{minipage}
  \Description{Twenty-window offline trajectory-rescheduling ablation for Tiny and Flash.
  Left panels show fewer rows under global rescheduling; Global FFD and \sysname{}
  have identical row counts. Right panels show that length-only scheduling
  raises the mean joint proxy, whereas \sysname{} reduces it in every optimizer window.}
  \caption{Offline trajectory rescheduling under shared expert placement: (a) Tiny; (b) Flash.
  Left: synchronized rows; right: per-window joint compute proxy relative to the
  DP-local FFD baseline.}
  \label{fig:offline-packing}
\end{figure*}

\paragraph{Compaction alone is insufficient.}
Global FFD needs fewer rows on every window of both models: the mean drops from
63.25 to 60.30 on Tiny and from 31.70 to 29.95 on Flash
(Table~\ref{tab:offline-packing}), i.e., mean per-window reductions of 4.65\%
and 5.64\%, with 4.91\% and 6.01\% more tokens per cell. Yet it raises the
joint proxy by 3.35\% on Tiny and 1.04\% on Flash and is worse on 20/20 and
17/20 windows, because higher EP peaks outweigh its lower dense-side cost.
Fewer, denser rows therefore do not by themselves imply a lower joint cost or
faster grouped-GEMM execution, which is why Global FFD serves online only as
the planner's fixed-row starting point, not as a deployed policy.

\paragraph{Routing-aware refinement.}
At the same global row count, \sysname{} lowers the joint proxy by 9.42\%
(Tiny) and 4.68\% (Flash) relative to Global FFD, accepting 4.96\% and 1.83\%
more dense-side cost in exchange for 21.77\% and 8.16\% less EP cost. Against
the in-house baseline, the final layout lowers the joint proxy on all 20
windows of each model: by 6.39\% on average for Tiny (range 3.44--8.62\%) and
3.69\% for Flash (2.65--4.87\%). Figure~\ref{fig:offline-packing} plots every
window rather than a selected example. These fixed-row gains come from seed
construction, cell pairing, and annealing together, not from annealing alone.

\paragraph{Checks and limitations.}
All 120 layouts keep exact trajectory coverage, expert demand, and the token
cap. The global row count reaches the capacity lower bound on 0/20 Tiny and
8/20 Flash windows; elsewhere it is a feasible heuristic count, not a certified
minimum. Re-running the baseline reconstruction with shuffle counters 0--4 in
addition to the window-ordinal counter gives 115 distinct window/counter
controls per model, and \sysname{}'s joint reduction stays positive in all of
them: 3.40--8.62\% on Tiny and 2.21--5.06\% on Flash. These are sensitivity
checks, not additional independent windows, and a single SA seed per window
does not establish robustness to the search seed. The study does not measure
kernel efficiency, pipeline bubbles, or end-to-end speedup.

\bibliographystyle{ACM-Reference-Format}
\bibliography{references}

\end{document}